\documentclass[english,aps,showpacs,showkeys,preprintnumbers,prd,floatfix,nofootinbib,superscriptaddress,notitlepage,hidelinks]{revtex4-1}

\usepackage{graphicx}
\usepackage{amsmath}
\usepackage{amsfonts}
\usepackage{hyperref}
\usepackage{color}
\usepackage{placeins}
\usepackage{tabu}
\usepackage{nicefrac}
\usepackage{ulem}
\usepackage{float}
\usepackage{subfig}
\usepackage{multirow}
\usepackage{tikz-feynman}
\newcommand{\orcid}[1]{\,\href{https://orcid.org/#1}{\includegraphics[width=9pt]{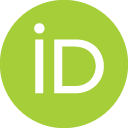}}\,}

\newcommand{\chidof}{\chi^2/N_{dof}}

\newcommand{\orcidPD}{0000-0001-7960-7953} %
\newcommand{\orcidHP}{0000-0001-8815-4255} %
\newcommand{\orcidVG}{0000-0002-2393-8507} %
\newcommand{\orcidIH}{0000-0003-1998-038X} %
\newcommand{\invgev}{\,GeV$^{-1}$}
\begin{document}
\title{Proton PDFs with non-linear corrections from gluon recombination}

\author{P.~Duwent\"aster\orcid{\orcidPD}}
\email{pitduwen@jyu.fi}
\affiliation{University of Jyvaskyla, Department of Physics, P.O. Box 35, FI-40014 University of Jyvaskyla, Finland}
\affiliation{Helsinki Institute of Physics, P.O. Box 64, FI-00014 University of Helsinki, Finland}

\author{V. Guzey\orcid{\orcidVG}}
\email{vadim.a.guzey@jyu.fi}
\affiliation{University of Jyvaskyla, Department of Physics, P.O. Box 35, FI-40014 University of Jyvaskyla, Finland}
\affiliation{Helsinki Institute of Physics, P.O. Box 64, FI-00014 University of Helsinki, Finland}

\author{I. Helenius\orcid{\orcidIH}}
\email{ilkka.m.helenius@jyu.fi}
\affiliation{University of Jyvaskyla, Department of Physics, P.O. Box 35, FI-40014 University of Jyvaskyla, Finland}
\affiliation{Helsinki Institute of Physics, P.O. Box 64, FI-00014 University of Helsinki, Finland}

\author{H. Paukkunen\orcid{\orcidHP}}
\email{hannu.t.paukkunen@jyu.fi}
\affiliation{University of Jyvaskyla, Department of Physics, P.O. Box 35, FI-40014 University of Jyvaskyla, Finland}
\affiliation{Helsinki Institute of Physics, P.O. Box 64, FI-00014 University of Helsinki, Finland}

\date{\today}
\begin{abstract}
\vspace*{0.5cm}

We present numerical studies of the leading non-linear corrections to the Dokshitzer-Gribov-Lipatov-Altarelli-Parisi evolution equations of parton distribution functions (PDFs) resulting from gluon recombination. The effect of these corrections is to reduce the pace of evolution at small momentum fractions $x$, while slightly increasing it at intermediate $x$. By implementing the non-linear evolution in the \textsc{xFitter} framework, we have carried out fits of proton PDFs using data on lepton-proton deep inelastic scattering from HERA, BCDMS and NMC. While we find no evidence for the presence of non-linearities, they cannot be entirely excluded either and we are able to set limits for their strength. In terms of the recombination scale $Q_r$, which plays a similar role as the saturation scale in the dipole picture of the proton, we find 
that $Q_r \lesssim 2.5 \, {\rm GeV}$. 
We also quantify the impact of the non-linear terms for the longitudinal structure function at the Electron-Ion Collider and the Large Hadron Electron
Collider and find that these measurements could provide stronger constraints on the projected effects.

\end{abstract}

\maketitle
\section{Introduction}
\label{sec:intro}

In the framework of Quantum Chromodynamics (QCD) and collinear factorization~\cite{Collins:1989gx}, the cross sections for hard processes involving intial-state hadrons are calculated as convolutions of perturbative, process-dependent coefficient functions and non-perturbative, process-independent parton distribution functions (PDFs) $f_i(x,Q^2)$. The PDFs describe the density of partons (gluons and quarks) of flavor $i$ carrying a fraction $x$ of the hadron's momentum at a resolution scale $Q$. While the $x$ dependence cannot be calculated by the perturbative means of QCD, its $Q^2$ dependence is governed by the Dokshitzer-Gribov-Lipatov-Altarelli-Parisi (DGLAP) evolution equations \cite{Dokshitzer:1977sg,Gribov:1972ri,Gribov:1972rt,Altarelli:1977zs}. These evolution equations originate from the renormalization of divergences induced by collinear radiation from the initial-state partons. As a result of the radiation, the density of partons that carry a momentum smaller than that from which the radiation tree started, 
grows, leading to a rise of PDFs at low momentum fractions $x$ as the scale $Q^2$ increases. This effect is especially pronounced for the gluon distribution. However, if the parton densities become sufficiently high, also the inverse processes in which several partons recombine, will begin to play a part, moderating the growth of PDFs at small $x$ as $Q^2$ increases. These contributions are formally of higher-twist, i.e., they are suppressed by inverse powers of $Q^2$. Since the $Q^2$-dependence of PDFs is only logarithmic, the effects of recombination 
should
disappear as $Q^2$ grows.
\\

In the language of PDFs, the first recombination terms in the evolution can be derived from diagrams, where two initial-state gluons merge into collinear gluons/quarks that eventually participate in the hard scattering. The leading contributions at small $x$ were originally calculated in Refs.~\cite{Qiu:1986wh, Kovchegov:2012mbw, Mueller:1985wy, Gribov:1983ivg} giving rise to the Gribov-Levin-Ryskin-Mueller-Qiu (GLR-MQ) corrections to the DGLAP evolution equations. Phenomenological applications of this approach have been studied in a number of publications, including its impact on observables in deep inelastic scattering (DIS) \cite{Bartels:1990zk}, comparisons with HERA data~\cite{Prytz:2001tb, Eskola:2002yc}, its relation to the Balitsky-Kovchegov (BK) evolution 
equation~\cite{Boroun:2013mgv, Boroun:2009zzb}, and the evolution of nuclear PDFs~\cite{Rausch:2022nkv}. However, the GLR-MQ equations violate the momentum conservation. This issue was fixed in the calculation of Zhu and Ruan~\cite{Zhu:1998hg,Zhu:1999ht}, which is based on the same class of diagrams, but keeps also the non-leading contributions, see also Ref.~\cite{Blumlein:2001bk}. The resulting evolution equation is the one whose implications we will here study the framework of global 
fits of PDFs. The effect of the non-linear corrections is to moderate the $Q^2$ growth of PDFs at small $x$, while somewhat increasing the speed of evolution at moderate values of $x$ such that the total momentum is conserved. This warrants to speak about dynamically generated shadowing and antishadowing -- terms which are perhaps more familiar from the context of nuclear PDFs \cite{Klasen:2023uqj}. 

For dimensional reasons, the higher-twist effects imply a presence of a dimensionful parameter $R$, which controls the strength of the non-linear effects and 
which 
can be associated with the size of the area in the transverse plane, where the partonic overlap leading to the non-linear corrections becomes important. The inverse of this parameter $1/R = Q_r$ gives correspondingly the momentum scale -- the recombination scale as we will call it -- below which the effects of recombination will be considerable. Once $Q^2$ is so low that the non-linear terms in the evolution are as important as the linear ones, also terms suppressed by higher inverse powers of $Q^2$ become presumably important and the resummation of all 
these terms
can be thought to give rise to the phenomenon of partonic saturation \cite{Gribov:1983ivg}. This regime is traditionally discussed in the framework of the Color Glass Condensate (CGC) effective field theory \cite{Gelis:2010nm,Morreale:2021pnn} and is implemented in the dipole picture of 
DIS.
In this case, the saturation effects are characterized by the saturation momentum scale $Q_s$, which can be related to the  critical dipole transverse size. It would be tempting to link the two emergent scales $Q_r$ and $Q_s$. However, in the dipole language $Q_s$ also depends on $x$ roughly as $Q_s \sim x^{-b}$, whereas in the picture discussed here it is a constant, with the $x$ dependence being entirely dynamical. Thus the two cannot be compared directly and the comparison is at best only indicative. 

In the presented work, we have implemented the first momentum-conserving non-linear corrections into the \textsc{HOPPET} evolution code~\cite{Salam:2008qg} and interfaced it with the \textsc{xFitter} framework~\cite{xFitter:2022zjb, Alekhin:2014irh}.  This allows us to perform global fits of proton PDFs with a range of different values for $R$. Scanning through various values of $R$ and inspecting the resulting goodnesses of the fits and comparisons with the data, we are able to place lower limits for $R$, or equivalently upper limits for the recombination scale $Q_r$. We have used BCDMS~\cite{BCDMS:1989qop}, HERA~\cite{H1:2015ubc} and NMC data~\cite{NewMuon:1996fwh} on lepton-proton DIS in our analysis. Apart from searching for signatures of and placing limits on the effects of recombination from these DIS data, our motivation is to qualitatively compare the systematics of non-linearities with those resulting from the Balitsky-Fadin-Kuraev-Lipatov (BFKL) resummation \cite{Fadin:1975cb,Kuraev:1976ge,Kuraev:1977fs,Balitsky:1978ic} of $\log x$-type terms relevant at small $x$. The BFKL resummation was incorporated into the DGLAP equations in Ref.~\cite{Bonvini:2017ogt} and evidence in favor of these dynamics has thereafter been seen in fits to the HERA data \cite{Ball:2017otu,xFitterDevelopersTeam:2018hym}. It could be expected that the general features of the partonic recombination and resumming BFKL logarithms are, however, qualitatively similar at small-$x$ as both tend to tame the rapid $Q^2$ growth of the gluon %
PDF.

While in the present paper we discuss the recombination in the case of free protons, we note that the non-linear corrections and partonic saturation are expected to be enhanced in the case of heavier nuclear targets \cite{Mueller:1985wy,Qiu:1986wh,Eskola:1993mb,Eskola:2003gc,Rausch:2022nkv}, the recombination scale increasing naively as $A^{1/6}$ as a function of the nuclear mass number $A$. As of today, however, no clear signal has been seen. To discover this enhancement is also one of 
the
cornerstones of the envisioned physics programs of the Electron-Ion Collider (EIC) \cite{Accardi:2012qut,AbdulKhalek:2021gbh} and Large %
Hadron Electron
Collider (LHeC) \cite{LHeC:2020van, LHeCStudyGroup:2012zhm} and a strong motivation to run the colliders with both light and heavy nuclei. The construction of a Future Circular Collider (FCC)~\cite{FCC:2018byv}, which would reach kinematics even further beyond those of the LHeC in electron-ion collisions, would naturally be able to put even stronger constraints on the nature and strength of non-linear corrections to the evolution of PDFs. 

The paper is organized as follows: First, Sec.~\ref{sec:theo} provides an overview of the theoretical background and the implementation of the Zhu and Ruan non-linear corrections to the evolution of PDFs, including numerical comparisons between the linear and non-linear evolution equations. In Sec.~\ref{sec:fit} we present our new global fits of proton PDFs with non-linear corrections. Section~\ref{sec:fl} discusses the impact on the longitudinal structure function $F_L(x,Q^2)$ in the HERA, EIC and LHeC kinematics. Finally, we summarize our findings and outline the way forward in Sec.~\ref{sec:conclusions}.

\section{Non-linear evolution equations}
\label{sec:theo}

\begin{figure}[hb]
    \centering
    \begin{tikzpicture}[baseline = 0cm]
    \begin{feynman}
        \vertex (proton1) {\(p\)};
        \vertex at ($(proton1) + (0cm, 0.1cm)$) (proton1up) [opacity = 0] {\(p\)};
        \vertex at ($(proton1) + (0cm, -0.1cm)$) (proton1down) [opacity = 0] {\(p\)};
        \vertex at ($(proton1) + (2cm, 0cm)$) (pBlob1);
        \vertex at ($(pBlob1) + (0cm, 0.1cm)$) (pBlob1up);
        \vertex at ($(pBlob1) + (0cm, -0.1cm)$) (pBlob1down);
        \vertex at ($(proton1) + (2.5cm, 0.5cm)$) (pBlobCircleLeft1);
        \vertex at ($(proton1) + (2.5cm, -0.5cm)$) (pBlobCircleLeft2);
        \vertex at ($(proton1) + (7.5cm, 0.5cm)$) (pBlobCircleRight1);
        \vertex at ($(proton1) + (7.5cm, -0.5cm)$) (pBlobCircleRight2);
        \vertex at ($(proton1) + (8cm, 0cm)$) (pBlob2);
        \vertex at ($(pBlob2) + (0cm, 0.1cm)$) (pBlob2up);
        \vertex at ($(pBlob2) + (0cm, -0.1cm)$) (pBlob2down);
        \vertex at ($(proton1) + (10cm, 0cm)$) (proton2) {\(p\)};
        \vertex at ($(proton2) + (0cm, 0.1cm)$) (proton2up) [opacity = 0] {\(p\)};
        \vertex at ($(proton2) + (0cm, -0.1cm)$) (proton2down) [opacity = 0] {\(p\)};
        \vertex at ($(pBlobCircleLeft1) + (0cm, 1cm)$) (ladderLeftBlueV1);
        \vertex at ($(pBlobCircleLeft1) + (0cm, 2cm)$) (ladderLeftBlueV2);
        \vertex at ($(pBlobCircleLeft1) + (0cm, 3.5cm)$) (ladderLeftBlueV3);
        \vertex at ($(pBlobCircleRight1) + (0cm, 1cm)$) (ladderRightBlueV1);
        \vertex at ($(pBlobCircleRight1) + (0cm, 2cm)$) (ladderRightBlueV2);
        \vertex at ($(pBlobCircleRight1) + (0cm, 3.5cm)$) (ladderRightBlueV3);
        \vertex at ($(ladderLeftBlueV3) + (0cm, 1cm)$) (box1TopLeft);
        \vertex at ($(ladderRightBlueV3) + (0cm, 1cm)$) (box1TopRight);
        \vertex at ($(box1TopLeft) + (0cm, 1cm)$) (box2BotLeft);
        \vertex at ($(box2BotLeft) + (0cm, 1cm)$) (box2TopLeft);
        \vertex at ($(box1TopRight) + (0cm, 1cm)$) (box2BotRight);
        \vertex at ($(box2BotRight) + (0cm, 1cm)$) (box2TopRight);
        \vertex at ($(box2TopLeft) + (-1cm, 1cm)$) (leptonGammaLeft);
        \vertex at ($(box2TopRight) + (1cm, 1cm)$) (leptonGammaRight);
        \vertex at ($(leptonGammaLeft) + (-1cm, 1cm)$) (leptonIn) {\(l\)};
        \vertex at ($(leptonGammaRight) + (1cm, 1cm)$) (leptonOut) {\(l\)};
        \vertex at ($(leptonIn) + (4.5cm, 0.5cm)$) (cutTop);
        \vertex at ($(cutTop) + (0cm, -11cm)$) (cutBot);
        \vertex at ($(pBlobCircleLeft1) + (1cm, 0cm)$) (ladderLeftRedV0);
        \vertex at ($(pBlobCircleLeft1) + (1cm, 1.5cm)$) (ladderLeftRedV1);
        \vertex at ($(pBlobCircleLeft1) + (1cm, 2.5cm)$) (ladderLeftRedV2);
        \vertex at ($(pBlobCircleLeft1) + (1cm, 3.5cm)$) (ladderLeftRedV3);
        \vertex at ($(pBlobCircleLeft1) + (4cm, 0cm)$) (ladderRightRedV0);
        \vertex at ($(pBlobCircleLeft1) + (4cm, 1.5cm)$) (ladderRightRedV1);
        \vertex at ($(pBlobCircleLeft1) + (4cm, 2.5cm)$) (ladderRightRedV2);
        \vertex at ($(pBlobCircleLeft1) + (4cm, 3.5cm)$) (ladderRightRedV3);

        \diagram* {
            (proton1) -- [fermion] (pBlob1),
            (proton1up) -- [fermion] (pBlob1up),
            (proton1down) -- [fermion] (pBlob1down),
            (pBlob1) -- [quarter left] (pBlobCircleLeft1),
            (pBlob1) -- [quarter right] (pBlobCircleLeft2),
            (pBlobCircleLeft1) -- (pBlobCircleRight1),
            (pBlobCircleLeft2) -- (pBlobCircleRight2),
            (pBlobCircleRight1) -- [quarter left] (pBlob2),
            (pBlobCircleRight2) -- [quarter right] (pBlob2),
            (pBlob2) -- [fermion] (proton2),
            (pBlob2up) -- [fermion] (proton2up),
            (pBlob2down) -- [fermion] (proton2down),
            (pBlobCircleLeft1) -- [gluon] (ladderLeftBlueV1),
            (ladderLeftBlueV1) -- [gluon] (ladderLeftBlueV2),
            (ladderLeftBlueV2) -- [gluon, edge label=\(x_1\)] (ladderLeftBlueV3),
            (pBlobCircleRight1) -- [gluon] (ladderRightBlueV1),
            (ladderRightBlueV1) -- [gluon] (ladderRightBlueV2),
            (ladderRightBlueV2) -- [gluon, edge label'=\(x_1\)] (ladderRightBlueV3),
            (ladderLeftBlueV1) -- [gluon] (ladderRightBlueV1),
            (ladderLeftBlueV2) -- [gluon] (ladderRightBlueV2),
            (ladderLeftBlueV3) -- (ladderRightBlueV3),
            (ladderLeftBlueV3) -- (box1TopLeft),
            (box1TopLeft) -- (box1TopRight),
            (ladderRightBlueV3) -- (box1TopRight),
            (box1TopLeft) -- [gluon, edge label=\(x\)] (box2BotLeft),
            (box1TopRight) -- [gluon, edge label'=\(x\)] (box2BotRight),
            (box2BotLeft) -- [fermion] (box2TopLeft),
            (box2TopLeft) -- [fermion] (box2TopRight),
            (box2TopRight) -- [fermion] (box2BotRight),
            (box2BotRight) -- [fermion] (box2BotLeft),
            (box2TopLeft) -- [boson, edge label=\(\gamma\)] (leptonGammaLeft),
            (box2TopRight) -- [boson, edge label'=\(\gamma\)] (leptonGammaRight),
            (leptonIn) -- [fermion] (leptonGammaLeft),
            (leptonGammaLeft) -- [fermion] (leptonGammaRight),
            (leptonGammaRight) -- [fermion] (leptonOut),
            (cutTop) -- [scalar] (cutBot),
            (ladderLeftRedV0) -- [red, gluon] (ladderLeftRedV1),
            (ladderLeftRedV1) -- [red, gluon] (ladderLeftRedV2),
            (ladderLeftRedV2) -- [red, gluon, edge label=\(x_1\)] (ladderLeftRedV3),
            (ladderLeftRedV1) -- [red, gluon] (ladderRightRedV1),
            (ladderLeftRedV2) -- [red, gluon] (ladderRightRedV2),
            (ladderRightRedV0) -- [red, gluon] (ladderRightRedV1),
            (ladderRightRedV1) -- [red, gluon] (ladderRightRedV2),
            (ladderRightRedV2) -- [red, gluon, edge label'=\(x_1\)] (ladderRightRedV3),
            
        };
        
    \end{feynman}
    \node[rectangle,draw, pattern=north west lines, pattern color=black, anchor=north west, minimum width=5cm,
                        minimum height = 1cm] (rect) at (box1TopLeft) {};
\end{tikzpicture}
    \caption{A typical cut diagram representing the recombination of two gluon ladders into one.
    Note that the gluon ladders in black and red do not interact. The momentum fractions $x$ and $x_1$ appearing in Eqs.~(\ref{eq:ZR_a}) and (\ref{eq:ZR_b}) are indicated.
    }
    \label{fig:ladderDiagram}
\end{figure}
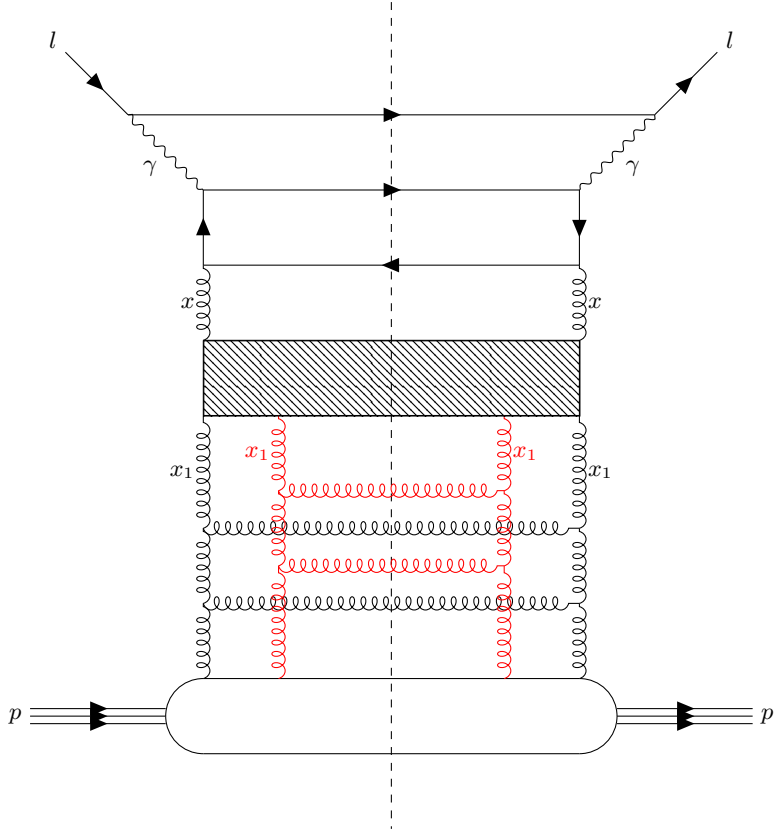

The DGLAP evolution equations are a set of renormalization group equations arising from the renormalization of collinear divergences in ladder-type Feynman graphs describing parton emission in QCD. They take the following standard form \cite{Dokshitzer:1977sg,Gribov:1972ri,Gribov:1972rt,Altarelli:1977zs}, 
\begin{align}
    Q^2\frac{d}{dQ^2}
    \begin{pmatrix}
        q_i(x,Q^2)\\[5pt]
        G(x,Q^2)
    \end{pmatrix}
    =\frac{\alpha_s(Q^2)}{2\pi} \sum_j \int_x^1\frac{dy}{y}
    \begin{pmatrix}
        P_{q_iq_j}\left(\frac{x}{y}\right) & P_{q_ig}\left(\frac{x}{y}\right)\\[5pt]
        P_{gq_j}\left(\frac{x}{y}\right) & P_{gg}\left(\frac{x}{y}\right)
    \end{pmatrix}
    \begin{pmatrix}
        q_j(y,Q^2)\\[5pt]
        G(y,Q^2)
    \end{pmatrix}\,,\label{eqn:DGLAP}
\end{align}
where $q_{i}(x,Q^2)$ are the PDFs for quark of flavor $i$, $G(x,Q^2)$ is the gluon PDF and $P_{ij}(x/y)$ are the partonic splitting functions. These equations are clearly linear in PDFs. The first non-linear corrections stem from diagrams like the one 
in Fig.~\ref{fig:ladderDiagram} showing a contribution to the DIS process with two gluons from the proton merging into one, which eventually splits into a quark-antiquark pair coupling to the virtual photon. Here, we will consider the corrections to the evolution of the gluon and quark-singlet distributions,

\begin{align}
\Sigma(x,Q^2)= \sum_i \left[q_i(x,Q^2)+\bar{q}_i(x,Q^2)\right], \ \ \mathrm{with} \ i\in\{u,d,s,c,b,t\}\,,
\end{align}
which can be written as \cite{Zhu:1999ht, Zhu:1998hg}, 
\begin{align}
    x\frac{dG(x,Q^2)}{d\ln{Q^2}}\ = & \quad \text{linear terms}\nonumber\\
        & + \frac{9}{32\pi^2}\left(\frac{1}{RQ}\right)^2\int_{x/2}^{1/2}dx_1xx_1G^2(x_1,Q^2)\sum_iP_i^{GG\rightarrow{G}}(x_1,x) \label{eq:ZR_a} \\
        & - \frac{9}{16\pi^2}\left(\frac{1}{RQ}\right)^2\int_{x}^{1/2}dx_1xx_1G^2(x_1,Q^2)\sum_iP_i^{GG\rightarrow{G}}(x_1,x),\nonumber\\
        \equiv & \quad \text{linear terms} + \Delta G(x, Q^2) \,, \nonumber\\
    x\frac{d\Sigma(x,Q^2)}{d\ln{Q^2}}\ = & \quad \text{linear terms}\nonumber\\
        & + \frac{9}{32\pi^2}\left(\frac{1}{RQ}\right)^2\int_{x/2}^{1/2}dx_1xx_1G^2(x_1,Q^2)\sum_iP_i^{GG\rightarrow{q\overline{q}}}(x_1,x) \label{eq:ZR_b}\\
        & - \frac{9}{16\pi^2}\left(\frac{1}{RQ}\right)^2\int_{x}^{1/2}dx_1xx_1G^2(x_1,Q^2)\sum_iP_i^{GG\rightarrow{q\overline{q}}}(x_1,x)\nonumber,\\
        \equiv & \quad \text{linear terms} + \Delta \Sigma(x, Q^2) 
        \,, \nonumber
\end{align}
where ``linear terms'' refer to the right-hand side of the usual DGLAP evolution, see Eq.~\eqref{eqn:DGLAP}. The square of the gluon PDF $G^2(x_1,Q^2)$ arises due to the modelling of the four-gluon correlation function, see the lower part of Fig.~\ref{fig:ladderDiagram}.
The parameter $R$ with the dimension of length is introduced on the basis of dimensional analysis. In a sense, the factor $1/(RQ)^2$ can be thought to be proportional to the overlap probability of two partons and $R$ regarded as the length scale on which it takes place. 
A smaller value for the parameter $R$ corresponds to stronger non-linear effects.
The recombination functions in Eqs.~\eqref{eq:ZR_a} and~\eqref{eq:ZR_b} are given by
\begin{align}
    \sum_i P_i^{GG\rightarrow{G}}(x_1,x) & = \frac{3\alpha_s^2}{8}\frac{C_A^2}{N^2-1}\frac{(2x_1-x)(-136xx_1^3-64x_1x^3+132x_1^2x^2+99x_1^4+16x^4)}{xx_1^5} \,,\\
    \sum_i P_i^{GG\rightarrow{q\overline{q}}}(x_1,x) & = \alpha_s^2\frac{T_f}{N(N^2-1)}(2x_1-x)^2\left[\frac{(4x^2+5x_1^2-6x_1x)}{x_1^5}+\frac{N^2}{(N^2-1)}\frac{(4x^2+4x_1^2-6x_1x)}{x_1^5}\right] \,,
\end{align}
with $C_A=N=3$ and $T_f=1/2$.
Unlike the GLR-MQ equations, these evolution equations conserve the PDF momentum sum rule,
\begin{equation}
\frac{d}{dQ^2} \int_0^1 dx x \left[ G(x,Q^2) + \Sigma(x,Q^2) \right] = 0 \,, 
\end{equation}
because the non-linear terms satisfy the following conditions separately for the gluon and singlet quark distributions,  
\begin{align}
    \frac{d}{d\ln{Q^2}}\int_0^1 dx\Delta G(x,Q^2)=0 \,,\\
    \frac{d}{d\ln{Q^2}}\int_0^1 dx\Delta \Sigma(x,Q^2)=0 \,.
\end{align}
Note that the linear parts of the full evolution equations also conserve the total momentum, but only in the sum of the gluon and quark singlet
PDFs. 
The method employed in Refs.~\cite{Zhu:1999ht, Zhu:1998hg} can also be used to derive the quark-antiquark and quark-gluon recombination, but these effects are omitted in the present analysis. We also note that the gluon and quark singlet 
PDFs
form a closed set of evolution equations, and that the quark flavor dependence enters through different non-singlet combinations of quarks, where the gluon recombination terms cancel. It is therefore consistent to consider the non-linear terms only in the evolution equations for gluon and quark singlet
PDFs, Eqs.~(\ref{eq:ZR_a}) and (\ref{eq:ZR_b}), while still solving for the full flavor-dependent evolution of PDFs. 

We have implemented the non-linear corrections as an extension to the evolution toolkit \textsc{HOPPET}~\cite{Salam:2008qg}. This code is optimized to rapidly solve the linear DGLAP equations by converting the numerical convolution of the PDFs with the splitting functions to multiplication of a matrix and a vector, which can be computed extremely fast on modern hardware.
To do this, HOPPET first changes the variables from $x$ and $Q^2$ to $y=\ln(1/x)$ and $t=\ln Q^2$, which helps with numerical accuracy. The PDF of each flavor can be represented numerically by the sum of a set of PDF values at fixed $y$ values $y_\alpha$, weighted with the interpolation weights $w_\alpha(y)$:
\\

\begin{align}
    q(y,t) \ \ &= \ \  \sum_\alpha w_\alpha(y) q(y_\alpha,t) \ \ \equiv \ \  \sum_\alpha w_\alpha(y)q_\alpha(t) \,.\label{eqn:hoppet1}\\
\intertext{Using Eq.~(\ref{eqn:hoppet1}), the convolution $(P\otimes q)$ can then be written in an analogous way,
}
    (P\otimes q)(y,t) \ \  &= \ \ \sum_\alpha w_\alpha(y)(P\otimes q)_\alpha(t) = \ \ \sum_{\beta}\sum_\alpha w_\alpha(y)P_{\alpha\beta}(t)q_\beta(t) \,,
    \intertext{where HOPPET precomputes the values of the matrix $P_{\alpha\beta}(t)$ once and reuses it for all future calculations,}
    P_{\alpha\beta}(t) \ \ &= \ \ \int_{e^{-y_\alpha}}^1 dz P(z,t)w_\beta(y_\alpha+\ln z) \,.
\end{align}
To apply this formalism to the non-linear terms, we treat $G^2(x,Q^2)$ as an independent flavor, which then enters the evolution as a linear term and is updated to be equal to $[G(x,Q^2)]^2$ at each iteration. The integration limits are handled by substituting $x$ with $x/2$ in the first non-linear term and by using a modified integration method, which integrates only up to $y=\ln(2)$ for the second term. This method allows us to compute the evolution with non-linear corrections, while increasing the computing time by no more than $\approx20\%$. We have also numerically checked our implementation by verifying the momentum sum rule.

\begin{figure*}[htb!]
	\centering
	\includegraphics[width=0.90\textwidth]{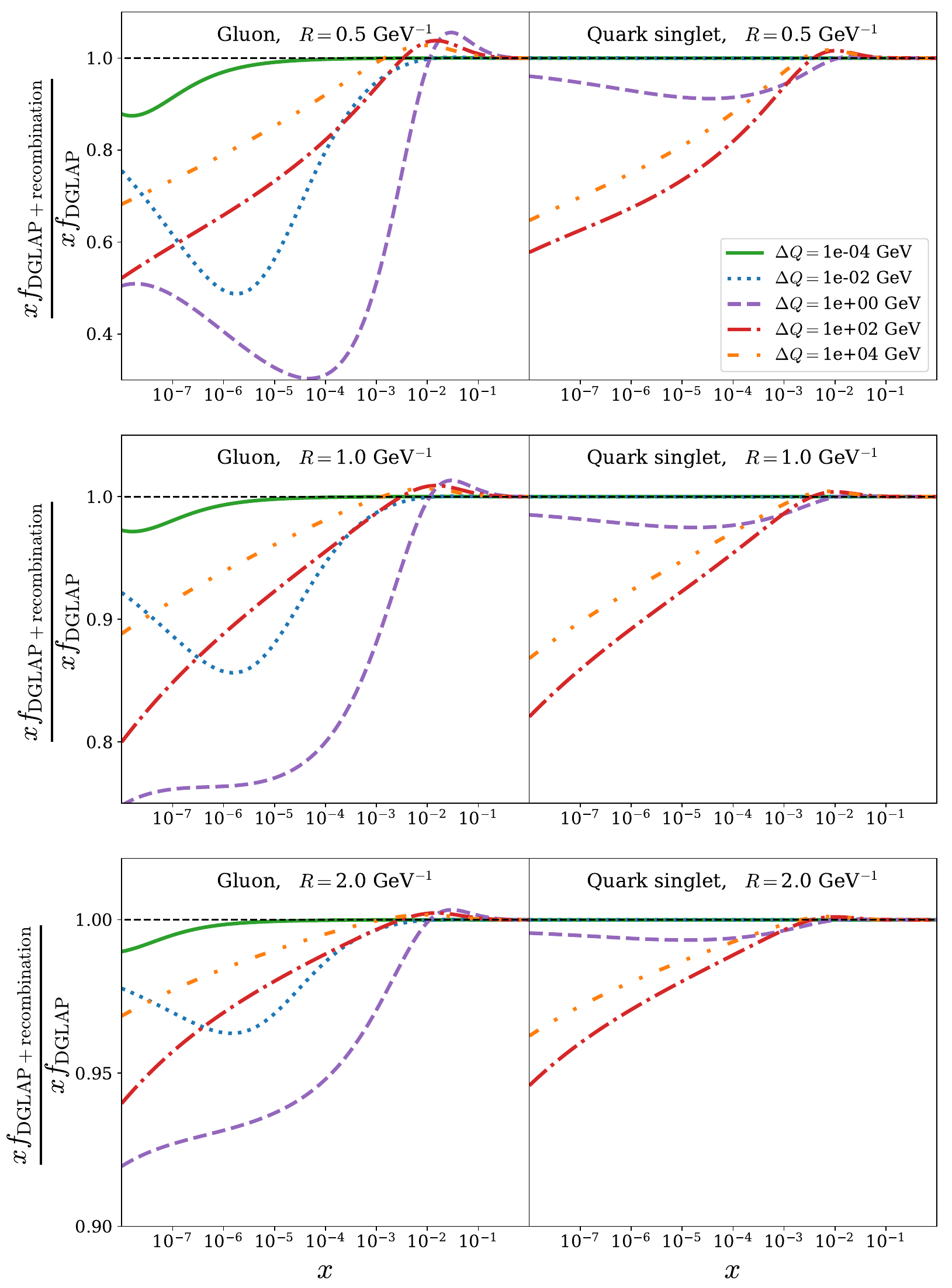}
	\caption{
  Ratios between %
  the PDFs evolved using the non-linear and linear evolution equations
  with the %
  CJ15 initial condition at $Q_0=1.3\,{\rm GeV}$
  as a function of $x$ and at different values of $Q=Q_0+\Delta Q $.
  The left column corresponds to the gluon
  PDF, 
  while the right column shows the quark singlet 
  distribution. 
  The top, middle and bottom rows correspond to $R=1/2$, 1 and 2\,GeV$^{-1}$, respectively.
 }
    \label{fig:qdep}
\end{figure*}

In Fig.~\ref{fig:qdep} we illustrate the effects of non-linearities in comparison to the regular DGLAP evolution 
and show the ratios between the PDFs evolved using the non-linear and
linear evolution equations
as a function of $x$ at different values of $Q$. Here, we have initialized the evolution with the boundary condition given by the CJ15 proton PDFs~\cite{Accardi:2016qay} at $Q_0 = 1.3$\,GeV, and performed the evolution in $Q^2$ according to the linear and non-linear equations up to different target values of $Q$. In both cases the linear terms in the evolution are included up to next-to-next-to-leading order (NNLO) accuracy in the strong coupling constant $\alpha_s$. The parameter $R$ is set to 0.5, 1 and 2\,GeV$^{-1}$ in the top, center and bottom panels, respectively -- we will see in the next section that these are reasonable choices. The $Q$ values are chosen as $Q_0 + \Delta Q$ with $Q_0 = 1.3$\,GeV and $\Delta Q$ increasing by factors of 100 from $10^{-4}$ to $10^4$\,GeV. The dynamically generated suppression at small $x$ (``shadowing'') and an enhancement at intermediate $x$ (``antishadowing'') are clearly visible for both 
quark singlet (right panels) and gluon (left panels) PDFs.
The effects are generally more significant for the gluon
PDF
than for the quark singlet, which, in part, results from the fact that the gluon distribution at the initial scale is smaller than the quark singlet 
one
-- very close to zero at small $x$ -- and therefore even small changes are more significant in the ratios. As a result, even a tiny step to higher $Q$ causes visible differences in the case of gluons, whereas for the quark singlet much larger step has to be taken in order to see a difference. While the relative effects of non-linearities first increase as $Q$ increases, there is eventually a turnaround and the differences begin to decrease. This was to be expected as $G^2(x,Q^2)/Q^2 \rightarrow 0$ as $Q^2 \rightarrow \infty$ so that the non-linear terms die away at asymptotically large $Q^2$. For gluons this turnaround happens already around $\Delta Q \approx 1\,{\rm GeV}$, but for quarks it takes place only at the electroweak scale $\Delta Q \approx 100\,{\rm GeV}$. The dependence on the choice of $R$ works as expected -- a smaller $R$ leads to more pronounced effects although qualitatively there are no significant differences between different choices of $R$.  
\\

\section{PDF fits with recombination effects}
\label{sec:fit}

As was observed in Fig.~\ref{fig:qdep}, the effects of recombination can persist up to rather high values of $Q^2$, where the PDFs are constrained by experimental data. As a result, in order to be consistent with these data constraints, the initial condition at $Q_0$ has to be iterated. To get a more complete and consistent picture of the impact of non-linearities it is thus necessary to perform global PDF fits including the non-linear terms in the evolution. In addition to the corrections to the evolution equations, higher-twist corrections also exist for the DIS coefficient functions like the ones for the structure function $F_2$ \cite{Zhu:1999ht, Zhu:1998hg,Blumlein:2001bk}. However, we neglect these effects in the present study.

\subsection{Fitting Framework}

The framework used to perform our global PDF fits with non-linear corrections is based on the \textsc{xFitter} package~\cite{xFitter:2022zjb, Alekhin:2014irh}, which we have extended by adding our modified version of \textsc{HOPPET} as a new module, see Sec.~\ref{sec:theo}. We have verified that this reproduces the results obtained with the default QCDNUM~\cite{Botje:2010ay, Botje:2016wbq} evolution in the limit $R\rightarrow \infty$. The fits are performed at the NNLO accuracy in both evolution and scattering coefficients. We use the RTOPT general-mass variable-flavor-number scheme \cite{Thorne:2006qt} for the heavy quark coefficient functions. In order to study the dependence of the non-linear effects on the assumed parameterization, we have prepared PDF fits with two different parameterizations for gluons and also varied the initial scale. We refer to these as parameterization 1 and 2.
For parameterization 1 the PDFs are parameterized at the initial scale $Q_0=1.0$\,GeV using the same parametric form as employed in the HERAPDF2.0 PDF analysis~\cite{H1:2015ubc}: For the valence distributions $u_v, d_v$, and the sea-quark densities $\bar{u}, \bar{d}$, we use the form
\begin{align}
    xf(x) = A_fx^{B_f}(1-x)^{C_f}\left(1+D_fx+E_fx^2\right) \,,
    \label{eqn:quarkparam}
\end{align}
and the gluon PDF is parameterized as
\begin{align}
    xg(x) = A_gx^{B_g}(1-x)^{C_g}-A^{'}_{g}x^{B^{'}_{g}}(1-x)^{C^{'}_{g}} \,.
    \label{eq:g_p1}
\end{align}
When using the parameterization in Eq.~(\ref{eq:g_p1}), the gluon distribution tends to become negative at small values of $x$ and $Q^2$. In the case of parameterization 2 we apply the same parameterization for gluons as we have for quarks in Eq.~(\ref{eqn:quarkparam}), which guarantees that the small-$x$ gluons will remain positive at the initial scale. To partly compensate for this restriction we increase the initial scale to $Q_0=1.3~\text{GeV}$ for parameterization 2. The strange quark is set to $xs = x\bar{s} = (2/3) x\bar{d}$ at $Q_0$ in both cases. 

The normalization parameters $A_{u_v}, A_{d_v}$ and $A_{g}$ are fixed through the quark-number and momentum sum rules. For simplicity and for lack of constraints for the flavor separation,
we fix $A_{\bar{u}} = A_{\bar{d}}$ and $B_{\bar{u}} = B_{\bar{d}}$.
With parameterization 1, we fix $C'_g=25$, where the exact value is unimportant as long as $C'_g\gg C_g$ such that the negative term does not contribute beyond low $x$. The full set of 14 open PDF parameters in the case of parameterization 1 is, 
\begin{align}
    \{B_g, C_g, A'_g, B'_g, B_{u_v}, C_{u_v}, E_{u_v}, B_{d_v}, C_{d_v}, C_{\bar{u}}, D_{\bar{u}}, A_{\bar{d}}, B_{\bar{d}}, C_{\bar{d}} \} \,,\label{eqn:parameters}
\end{align}
and in the case of parameterization 2,
\begin{align}
    \{B_g, C_g, D_g, E_g, B_{u_v}, C_{u_v}, E_{u_v}, B_{d_v}, C_{d_v}, C_{\bar{u}}, D_{\bar{u}}, A_{\bar{d}}, B_{\bar{d}}, C_{\bar{d}} \} \,. \label{eqn:parameters2}
\end{align}
Any parameters not explicitly mentioned are set to zero. The parameter $R$ is kept fixed in each fit and the procedure is repeated for $R$ values from 0.2\,GeV$^{-1}$ to 3.0\,GeV$^{-1}$ in steps of 0.01\invgev. For $R\gtrsim 3.0$\,GeV$^{-1}$, the non-linear modifications were found to be negligible. 

We find the optimal set of parameters by minimizing the $\chi^2$ function defined as
\begin{align}
    \chi^2 = \sum_i \frac{[T_i - \sum_\alpha \gamma_\alpha^iD_ib_\alpha-D_i]^2}{\left(\delta_{i,\mathrm{stat}}\sqrt{D_iT_i}\right)^2+\left(\delta_{i,\mathrm{uncorr}}T_i\right)^2} + \sum_\alpha b_\alpha^2.
\end{align}
The theoretical predictions for data point $i$ and a given set of PDF parameters are represented by $T_i$. The measured data values are given by $D_i$ and $\delta_{i,\mathrm{stat}}, \delta_{i,\mathrm{uncorr}}$ and $\gamma_\alpha^i$ are the statistical, uncorrelated and correlated uncertainties associated with each data point. 
Minimizing this $\chi^2$ function with respect to the 14 fit parameters and the nuisance parameters $b_\alpha$ determines the central set of PDFs and systematic shifts, $- \sum_\alpha \gamma_\alpha^iD_ib_\alpha$, which give the best description of the data. The uncertainties of the obtained PDFs are then determined using the Hessian method~\cite{Pumplin:2001ct}, which is based on a quadratic expansion of the $\chi^2$ function around its minimum, $\chi^2 = \chi^2_{\mathrm{min}}+\Delta \chi^2$. The choice of the maximum displacement $\Delta \chi^2$, the tolerance, depends on the definition of uncertainties and in PDF fits it is usually conservatively chosen to be $\Delta \chi^2 \gg 1$. Indeed, assuming that the data points are Gaussianly distributed around the "truth", the expected $\chidof$ should be less than $\sim$$1.05$ in 90\% of the cases with $\sim$$1600$ data points, which is the amount of points we are now considering. However, the minimum values we find are $\chidof \sim$$1.18$, i.e., the ideal choice $\Delta \chi^2 = 1$ is, perhaps, an underestimate. Assuming a Gaussian probability density for the fit parameters results in $\Delta \chi^2_{\mathrm{max}} \approx 20$ at the 90\% confidence level \cite{Eskola:2009uj}. 
In practice, we determine the PDFs with the tolerance set to $\Delta \chi^2 = 1$ and then rescale the resulting uncertainties by $\sqrt{20} \approx 4.5$ to match the actual tolerance since this leads to a better convergence of the \textsc{xFitter} algorithms.

\subsection{Data selection}

Our analysis, as the ones presented in Refs.~\cite{Walt:2019slu} and \cite{Helenius:2021tof}, includes the lepton-proton DIS data from the BCDMS, HERA and NMC experiments, 
which are 
summarized in Table~\ref{tab:data}. The data from BCDMS and NMC are given in terms of the structure function $F_2^p(x,Q^2)$ measured in neutral-current (NC) deep inelastic muon scattering. The HERA data sets provide the reduced cross section of DIS in electron-proton and positron-proton collisions, both for charged-current (CC) and NC processes. We do not include the HERA data on the longitudinal structure function $F_L$~\cite{H1:2013ktq} since it is derived from the other HERA data sets, and including it would to some extent double-count some constraints. We will, however, discuss $F_L$ separately in Sec.~\ref{sec:fl}. The data on heavy-quark production from HERA are omitted as well \cite{H1:2009uwa,H1:2012xnw,ZEUS:2014wft}. 

In the first round of fits, we apply the same $Q^2>3.5$\,GeV$^2$ cut as, e.g., in the HERAPDF2.0 analysis, but with parameterization 1 we also perform a second round of fits with the cut lowered to $Q^2>1.0$\,GeV$^2$
to see whether the inclusion of the $1/Q^2$ suppressed terms in the evolution equation can provide an improved description of the data in the low-$Q^2$ region. Indeed, it is known that in NNLO fits, 
$\chidof$ tends to get systematically worse as the minimum $Q^2$ is lowered from $15\,{\rm GeV}^2$ to $2\,{\rm GeV}^2$ \cite{Ball:2017otu,xFitterDevelopersTeam:2018hym} -- a tension 
that
the BFKL resummation appears to ease. The two rightmost columns of Table~\ref{tab:data} indicate the number of data points per experiment, which pass these cuts. The total number of data points is 1568 for the higher-$Q^2$ cut and 1636 for the lower-$Q^2$ one, respectively.

\begin{table*}[htbp!]
    \renewcommand{\arraystretch}{1.4}
    \setlength\tabcolsep{4pt}
	\caption{Summary of DIS data sets used in out fits of proton PDFs with non-linear corrections. The last two columns show the number of data points passing the $Q^2>3.5$\,GeV$^2$ and $Q^2>1$\,GeV$^2$ cuts, respectively. The energies listed for the HERA measurements refer to the proton beam energy, while those listed for BCDMS refer to the muon beam energy. NMC uses the same energies as BCDMS but combines them into one dataset.}
	\centering	
	\begin{tabular}{|c|c|c|c|c|c|c|}
		\hline 
		Experiment & Data set & Year & Ref. & $N_{\text{data}}^{Q^2>3.5\,\mathrm{GeV^2}}$ & $N_{\text{data}}^{Q^2>1.0\,\mathrm{GeV^2}}$ \\ 
		\hline
		\hline
		\multirow{4}{*}{BCDMS}  & NC $\mu$ $F_2^p$ 100\,GeV & \multirow{4}{*}{1996}  & \multirow{4}{*}{\cite{BCDMS:1989qop}}  & 83 & 83  \\ 
                                                        & NC $\mu$ $F_2^p$ 120\,GeV &  & & 91 & 91  \\ 
                                                        & NC $\mu$ $F_2^p$ 200\,GeV & &  & 79 & 79  \\ 
                                                        & NC $\mu$ $F_2^p$ 280\,GeV & &  & 75 & 75  \\ 
		\hline 
		
        \multirow{7}{*}{HERA} & NC e$^+$ 920\,GeV & \multirow{7}{*}{2015}  & \multirow{7}{*}{\cite{H1:2015ubc}}  & 377 & 425  \\ 
                                                        & NC e$^+$ 820\,GeV & & & 70 & 78  \\ 
                                                        & NC e$^+$ 575\,GeV & & & 254 & 260  \\ 
                                                        & NC e$^+$ 460\,GeV & & & 204 & 210  \\ 
                                                        & NC e$^-$ 920\,GeV                  & & & 159 & 159  \\ 
                                                        & CC e$^+$ 920\,GeV                    & & & 39 & 39  \\ 
                                                        & CC e$^-$ 920\,GeV                   & & & 42 & 42  \\ 
		\hline 
		NMC  & NC $\mu$ $F_2^p$ & 1997 & \cite{NewMuon:1996fwh} & 95 & 95  \\ 
		\hline 
            \hline
            \multicolumn{4}{|c|}{Total} & 1568 & 1636 \\
            \hline
	\end{tabular}     	
	\label{tab:data}
\end{table*}

\subsection{Fit results}

\begin{figure}
    \centering
    \includegraphics[width=0.7\textwidth]{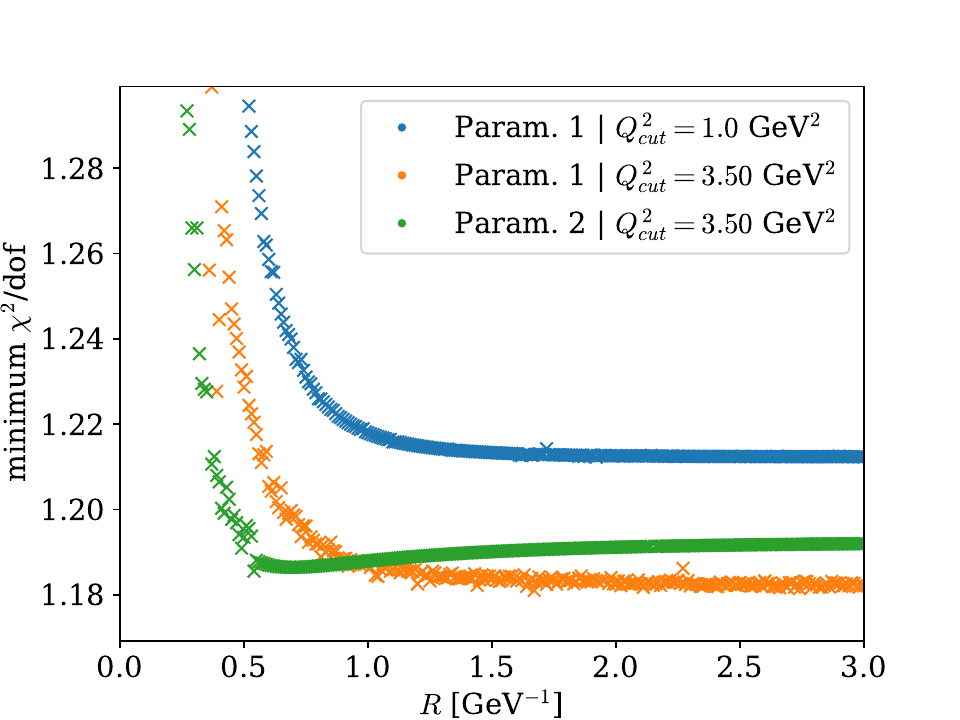}
    \caption{Minimum $\chidof$ as a function of $R$ for parameterization 1 with $Q^2 > 3.5$\,GeV$^2$  (orange) and $Q^2 > 1$\,GeV$^2$ (blue), and
    parameterization 2 with $Q^2 > 3.5$\,GeV$^2$ (green). 
    }
    \label{fig:Chi2Minima}
\end{figure}

The resulting values of $\chidof$ minima for all the fits with different $R$ values are shown in Fig.~\ref{fig:Chi2Minima}
and the contributions of different data sets to the total $\chidof$ are presented in 
Appendix \ref{sec:chi2data}, Tables~\ref{tab:chi2table}, \ref{tab:chi2table_b} and \ref{tab:chi2table_c}.
In Fig.~\ref{fig:Chi2Minima}, the orange (parameterization 1) and green (parameterization 2) crosses correspond to the $Q^2 > 3.5$\,GeV$^2$ cut
and the blue ones (parameterization 1) are for the fits with $Q^2>1.0$\,GeV$^2$. 
We have not performed a fit with parameterization 2 with the lower cut, as it lies below the initial scale of the PDF evolution in this case. 

In the case of parameterization 1, the resulting values of $\chidof$ generally decrease with increasing $R$, i.e., with a decreasing strength of the non-linear corrections, but remain constant for $R>1.5$\,GeV$^{-1}$. Lowering the value of $R$ down to $\approx0.7$\,GeV$^{-1}$ causes only a small increase, but below that 
$\chidof$ rapidly starts diverging. Comparing the curves between the two different cuts, there is no qualitative difference in the $R$ dependence, but the lower cut shifts $\chidof$ upwards by 
approximately 0.04 units
across all values of $R$. While this increase in $\chidof$ from the lowered cut might not seem too significant at first sight, one has to consider the small number of added data points that causes it. Namely, the 68 added data points increase the total $\chi^2$ by 145, i.e., more than 2 units per point. 
Therefore, the non-linear corrections do not help to alleviate the tensions with lower-$Q^2$ data. 
In the case of parametrization 2 (green crosses), i.e., with a positive-defined gluon distribution at the initial scale, the best descriptions are obtained in the interval $0.5 < R < 0.7$\,GeV$^{-1}$ although the difference with respect to $R=3$\,GeV$^{-1}$ is small. 

Tables~\ref{tab:chi2table}, \ref{tab:chi2table_b} and \ref{tab:chi2table_c} in Appendix \ref{sec:chi2data} break down the obtained $\chi^2$ into contributions from different data sets. Tables~\ref{tab:chi2table} and \ref{tab:chi2table_c}, which correspond to the fits with parameterization 1 at $Q^2_{\mathrm{cut}}=3.5$\,GeV$^2$ and 1\,GeV$^2$, respectively, show that the preference towards higher values of $R$ is a feature generally shared by all the data sets. At the same time, some data sets are more sensitive to $R$ than others with the NMC data set showing the largest difference in $\chidof$. Table \ref{tab:chi2table_b} shows that 
with parametrization 2, the $\chidof$ profiles as a function of $R$ are flatter than in the case of parametrization 1.

\begin{figure}[t]
    \centering
    \includegraphics[width=0.8\textwidth]{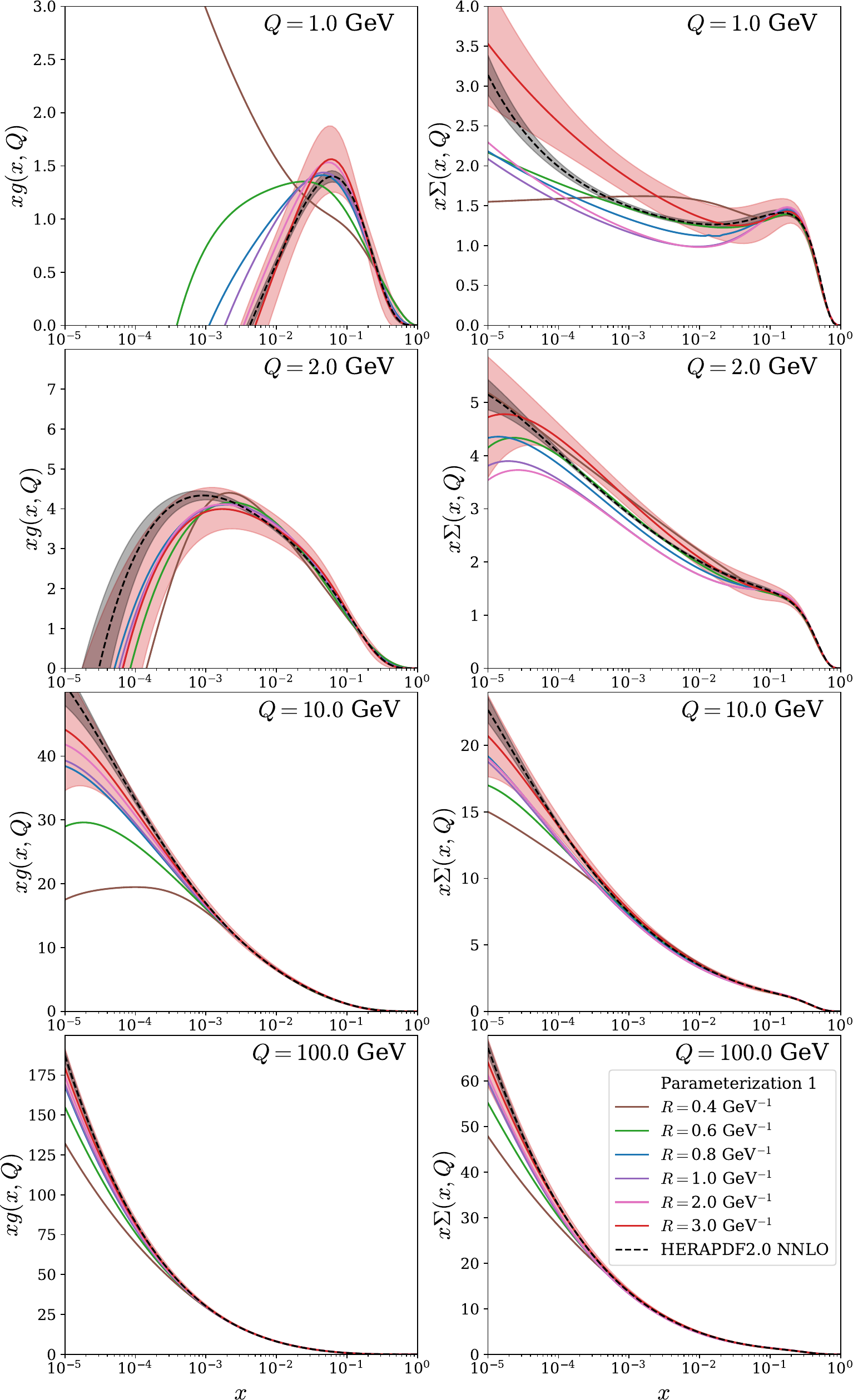}
    \caption{PDFs resulting from fits with parameterization 1 at various $R$ values. The left column shows the gluon PDF, while the right side shows the quark singlet PDF. The rows correspond to different scales $Q$. %
    }
    \label{fig:PDFcomp}
\end{figure}

\begin{figure}
    \centering
    \includegraphics[width=0.8\textwidth]{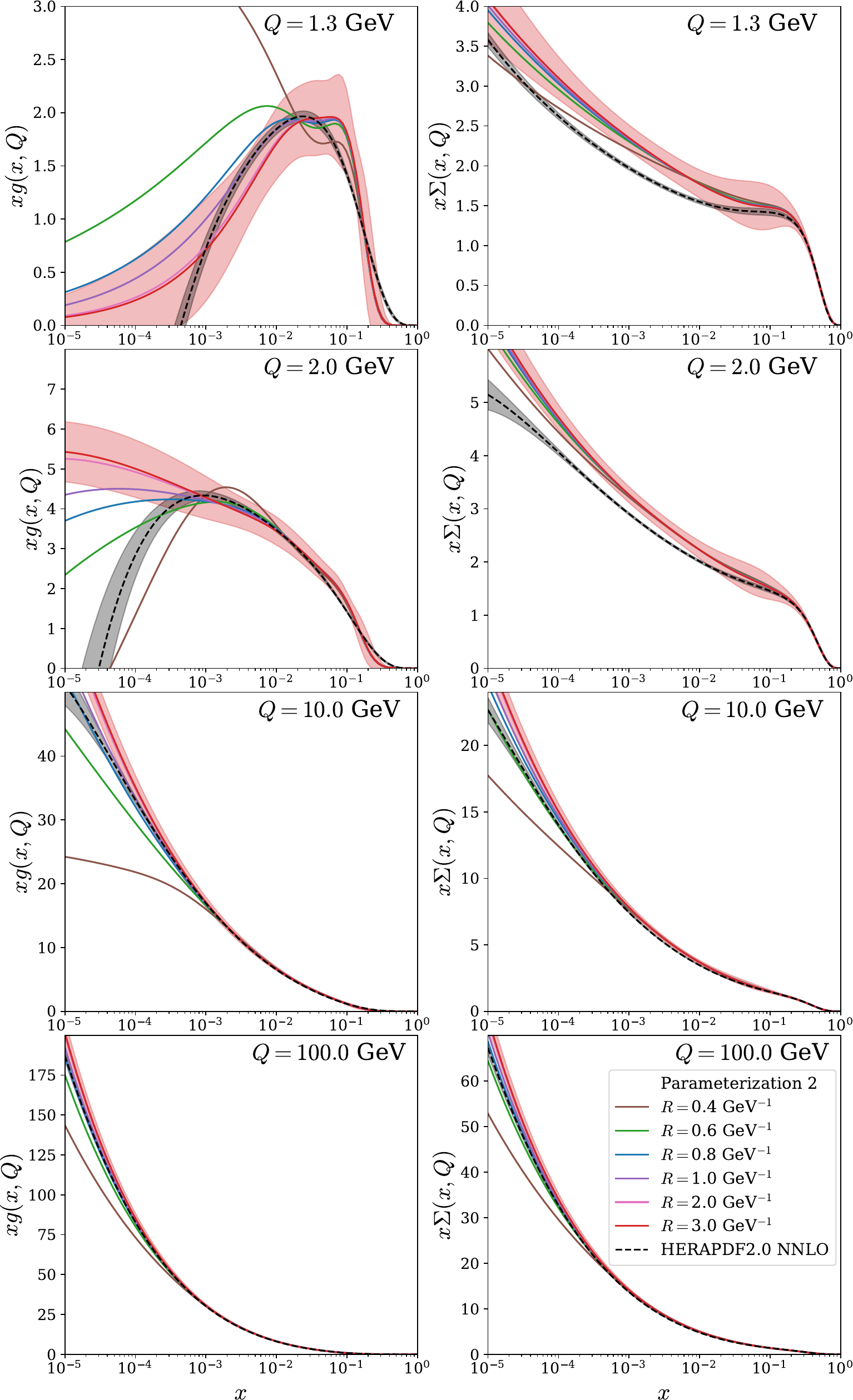} 
    \caption{The same as Fig.~\ref{fig:PDFcomp}, but showing PDFs fitted with parameterization 2.}
    \label{fig:PDFcomp_b}
\end{figure}

In the following, we will discuss proton PDFs determined from the fits with the $Q^2>3.5$\,GeV$^2$ cut for a representative selection of $R$ values of $\{0.4, 0.6, 0.8, 1, 2, 3\}$\,GeV$^{-1}$. Figures~\ref{fig:PDFcomp} and \ref{fig:PDFcomp_b} show the resulting gluon (left) and quark singlet (right) distributions at $Q=\{Q_0, 2\,{\rm GeV} , 10\,{\rm GeV}, 100\,{\rm GeV}\}$ for parameterization 1 and 2, respectively. They are compared to the HERAPDF2.0 results given by the black dashed curve. To avoid visual clutter, the uncertainties (90\% confidence level) are only shown for HERAPDF2.0 and $R=3.0$\,GeV$^{-1}$ case, but the remaining uncertainties of the rest closely resemble the latter one. The 90\% confidence level for HERAPDF2.0 is obtained by multiplying the uncertainties, which are given at 68\% confidence level with a factor of 1.645.

In the case of parameterization 1, the fit with the highest value of $R$, i.e., the one with smallest non-linear effects, behaves similarly as the HERAPDF2.0 reference but, despite the added data from BCDMS and NMC, has significantly larger uncertainties due to the higher Hessian tolerance -- 20 for our fit vs.~2.71 for HERAPDF2.0 at 90\% confidence level. 
The differences in the central values can, at least partly, be attributed to the fact the we included these additional data, which may affect the PDFs even at low-$x$ due to the sum rules.
As the value of $R$ is lowered, the shapes of the fitted PDFs change progressively further away from the baseline with $R=3.0$\,GeV$^{-1}$. In particular, one observes a significant suppression of the gluon and quark-singlet PDFs at small $x$ and $Q\ge 2$\,GeV as $R$ decreases. 
This suppression becomes weaker as the scale $Q$ is increased due to the weakening of the non-linear effects by a factor of $1/Q^2$ in the evolution equations, but it is still visible even at $Q=100$\,GeV. As expected, the largest differences between the PDFs corresponding to different values of $R$ are visible at the initial scale $Q_0=1$\,GeV, though they carry little meaning since there are no data that constrain this region due to the $Q^2_\mathrm{cut}>3.5$\,GeV$^2$ cut.

A comparison of the results presented Figs.~\ref{fig:PDFcomp} and \ref{fig:PDFcomp_b} shows that the systematics of the $R$ and $Q$ dependence of the PDFs obtained using parameterizations 1 and 2 are very similar even though there are large differences around the initial scale $Q_0$. This gives us confidence that our conclusions are not very sensitive to the details of the parameterization. 
The parameters of the fits are summarized in Tables~\ref{tab:paramtable}, \ref{tab:paramtable_b} and \ref{tab:paramtable_c} in Appendix \ref{sec:pdfparams}.

\section{Impact on the longitudinal structure function }\label{sec:fl}

The longitudinal structure function $F_L$ carries a direct sensitivity to the gluon PDFs and should therefore have an increased sensitivity to the non-linear effects.  
We note that since the HERA data entered our analysis as reduced cross sections $\sigma_{r} = F_2 - (y^2/Y_+)F_L$, where $Y_+ = 1+(1-y)^2$, $y=s/xQ^2$, $s$ being the squared center-of-mass energy of the lepton-proton collision, they already carry information of $F_L$. In addition, the $Q^2$ dependence of $F_2$ is directly sensitive to $F_L$ \cite{Lappi:2023lmi}, particularly at small $x$. In this sense the discussion below is not entirely independent of the fits presented above. 

\subsection{The longitudinal structure function at HERA}

\begin{figure}
    \centering
    \includegraphics[width=0.48\textwidth]{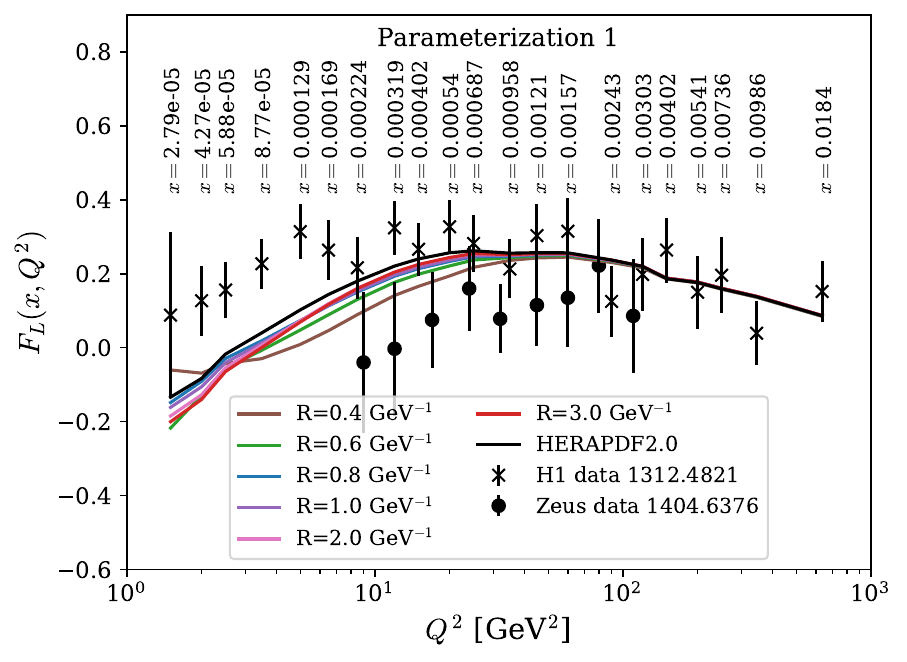}
    \includegraphics[width=0.48\textwidth]{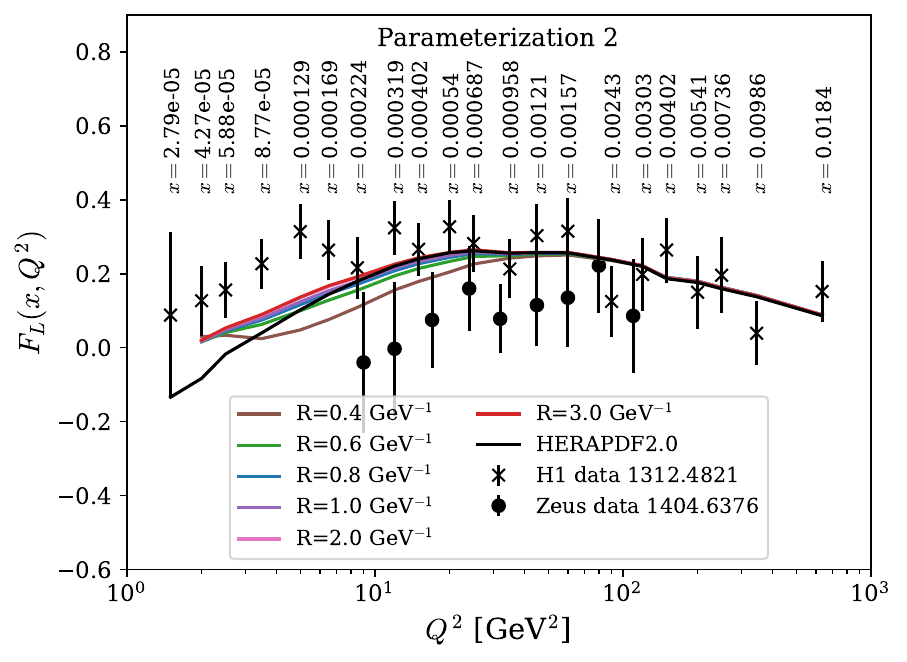}
    \caption{Structure function $F_L$ calculated with PDFs resulting from different choices of $R$ compared with the measurements by H1~\cite{H1:2013ktq} and ZEUS~\cite{ZEUS:2014thn}.  We only show predictions for $Q^2>Q_0^2$.
    }
    \label{fig:HERA_FLQ1}
\end{figure}

\begin{figure}
    \centering
    \includegraphics[width=0.98\textwidth]{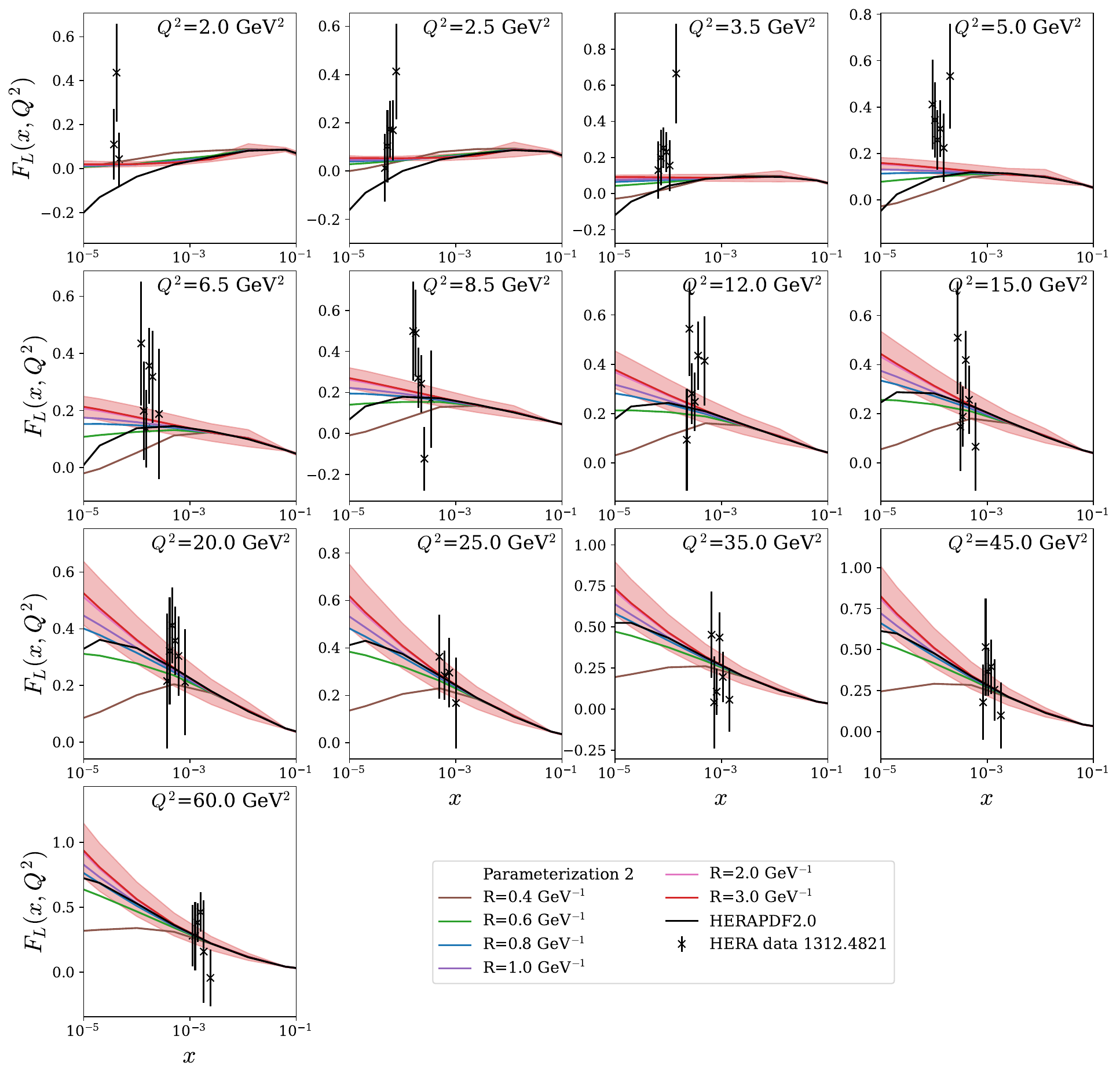}
    \caption{Structure function $F_L$ resulting from
    PDFs fitted with parameterization 2 and different $R$ values compared with the HERA $F_L$  measurements~\cite{H1:2013ktq}. We only show predictions for $Q^2>Q_0^2$.}
    \label{fig:HERA_FL_b}
\end{figure}

To gain a deeper understanding of the sensitivity of the HERA data to non-linear corrections, we compare the values for the longitudinal structure function $F_L(x,Q^2)$ that result from our PDFs  
with the data taken by H1~\cite{H1:2013ktq} and ZEUS~\cite{ZEUS:2014thn}. 
Figure~\ref{fig:HERA_FLQ1} presents the HERA $F_L$ measurements at different values of $Q^2$ and integrated over $x$ bins. The left panel shows the data compared with the predictions using the PDFs fitted at $Q^2_{\mathrm{cut}}=3.5$\,GeV$^2$ with parameterization 1 and the right panel with parameterization 2. For $Q^2\gtrsim100$\,GeV$^2$ the predictions converge and become indistinguishable. Towards lower $Q^2$ values the predictions spread out with higher $R$ values corresponding to slightly larger values of $F_L$. At low $Q^2$, the behaviour depends on the parameterization. In the case of parameterization 2, the predictions converge again towards $F_L\rightarrow 0$ at the PDFs parameterization scale $Q_0^2=1.69$\,GeV$^2$. %
The predictions for parameterization 1 do not converge towards $F_L\rightarrow 0$ since the parameterization allows the PDFs to be negative and therefore yield negative $F_L$ values for $Q^2\lesssim 5$\,GeV$^2$. 
At $Q^2\lesssim10$\,GeV$^2$, the NNLO predictions generally lie below the values for $F_L$ measured by H1, as has been observed in other studies~\cite{Ball:2017otu, xFitterDevelopersTeam:2018hym}. 
On the other hand, the ZEUS measurements generally lie below the predictions, but come with larger uncertainties.
Interestingly, the systematics of non-linearities we find goes in the oppostite direction in comparison to the effects of small-$x$ BFKL resummation: While the non-linear effects tend to decrease $F_L$, the small-$x$ resummation has a tendency to increase it. This can be traced to the small-$x$ behaviour of gluon and quark-singlet PDFs which are, at perturbative values of $Q^2$, increased in the BFKL approach in comparison to the fixed-order NNLO result. In the case of non-linearities discussed here, the gluon and quark-singlet PDFs are generally below the standard NNLO result at small $x$. At very low values of $Q^2$ -- very close to the parametrization scale -- they go in the opposite way, but the evolution is slower in the case with non-linear effects and leads to suppressed gluon and quark singlet PDFs very quickly as $Q^2$ increases. As a result, the systematics of BFKL and non-linear effects are quite different. 
Figure~\ref{fig:HERA_FL_b} shows $F_L$ as a function of $x$ at different values of $Q^2$. The theoretical calculations are performed with the PDFs corresponding to fits with different values of $R$ using parameterization 2. Predictions for $F_L$ made with PDFs from parameterization 1 result in negative values of $F_L$ at $Q^2 \leq 10$\,GeV$^2$ due to the gluon PDFs not being restrained to positive values at the initial scale. Since the negativity of $F_L$ seen in the previous figure indicates that the corresponding theoretical setup is not entirely consistent, e.g., too low $Q^2$ or too low $R$, we omit the PDFs fitted with parameterization 1 from the following discussion.
For clarity, we show the uncertainty band only for the $R=3$\,GeV$^{-1}$ case. One can see from the figure that for $x\gtrsim0.3\times10^{-2}$, the predictions with different $R$ are almost indistinguishable. However for $Q^2>5$\,GeV$^2$ and towards lower $x$ values, the calculations with a smaller $R$ lead to significantly reduced $F_L$. This was to be expected as both gluon and quark-singlet PDFs at small $x$ were found to be increasingly suppressed as $R$ decreased, see Figs.~\ref{fig:PDFcomp} and \ref{fig:PDFcomp_b}. At lower values of $Q^2$, where the relative differences between predictions corresponding to different $R$ values are expected to be the largest, the absolute values are too close to zero to see any meaningful differences. As already observed in Sec.~\ref{sec:theo}, the effects of non-linearities die out rather slowly at small $x$ as $Q^2$ increases. Also here, we see in Fig.~\ref{fig:HERA_FL_b} that the differences between calculations with different $R$ actually still increase at $x \approx 10^{-5}$ until the highest considered $Q^2$. We note that at $Q^2>8.5$\,GeV$^2$ the $F_L$ data are well described by most of the predictions. However, with $R\leq 0.4$\,GeV$^{-1}$ the description begins to visibly deteriorate even at higher values of $Q^2$.

\subsection{Prospects for the longitudinal structure function at EIC and LHeC}

\begin{figure}
    \centering
    \includegraphics[width=0.98\textwidth]{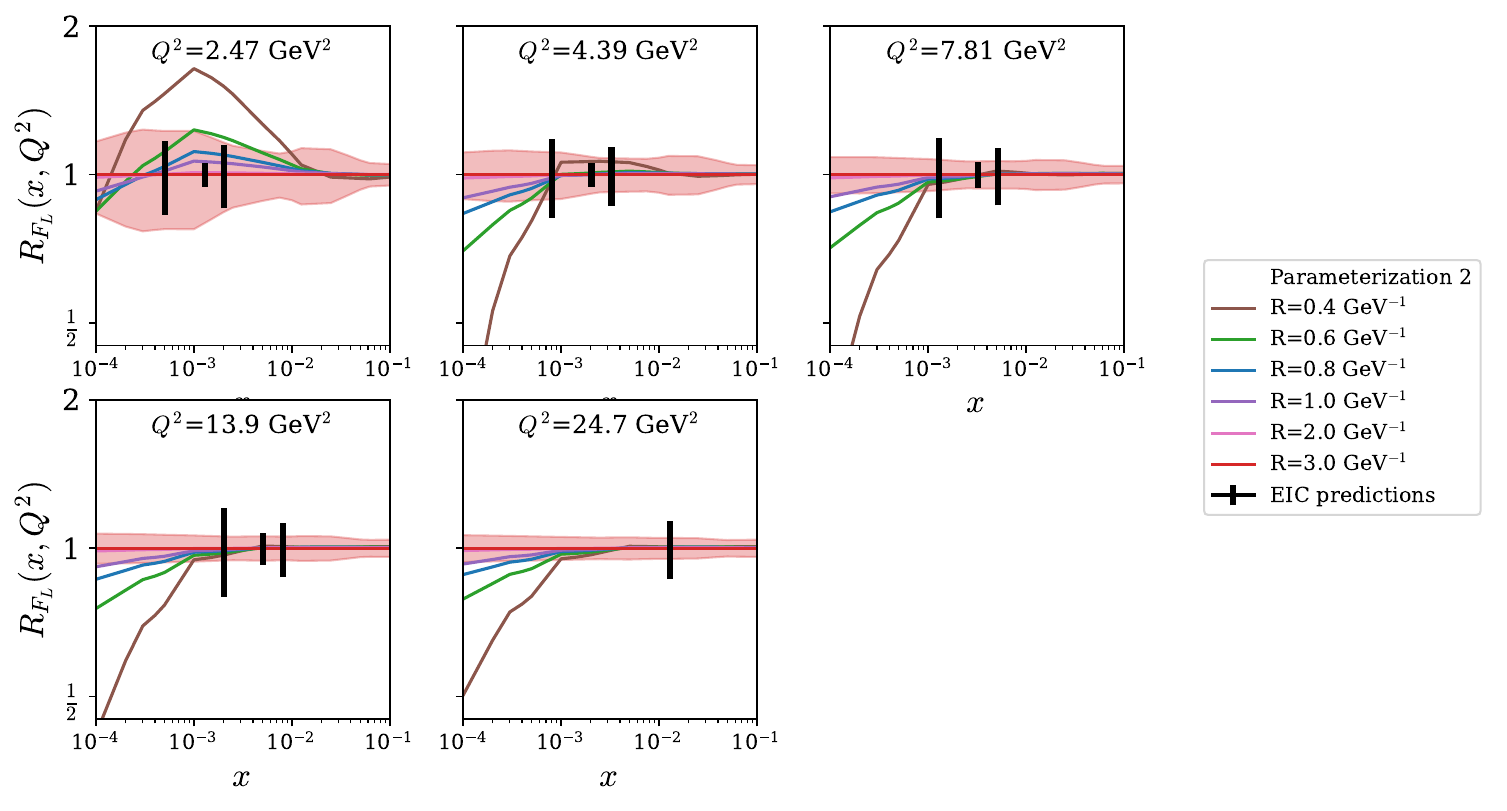}
    \caption{The ratios of the longitudinal structure functions corresponding to different values of $R$ to that with $R=3$\,GeV$^{-1}$, see Eq.~(\ref{eqn:rfl}), as a function of $x$. The calculations use parametrization 2 and are carried out at different values of $Q^2$.
    The vertical bars represent the expected statistical uncertainty of the corresponding measurements at the EIC~\cite{Aschenauer:9999abc}.  We only show predictions for $Q^2>Q_0^2$.}
    \label{fig:EIC_FL_2}
\end{figure}

\begin{figure}
    \centering
    \includegraphics[width=0.98\textwidth]{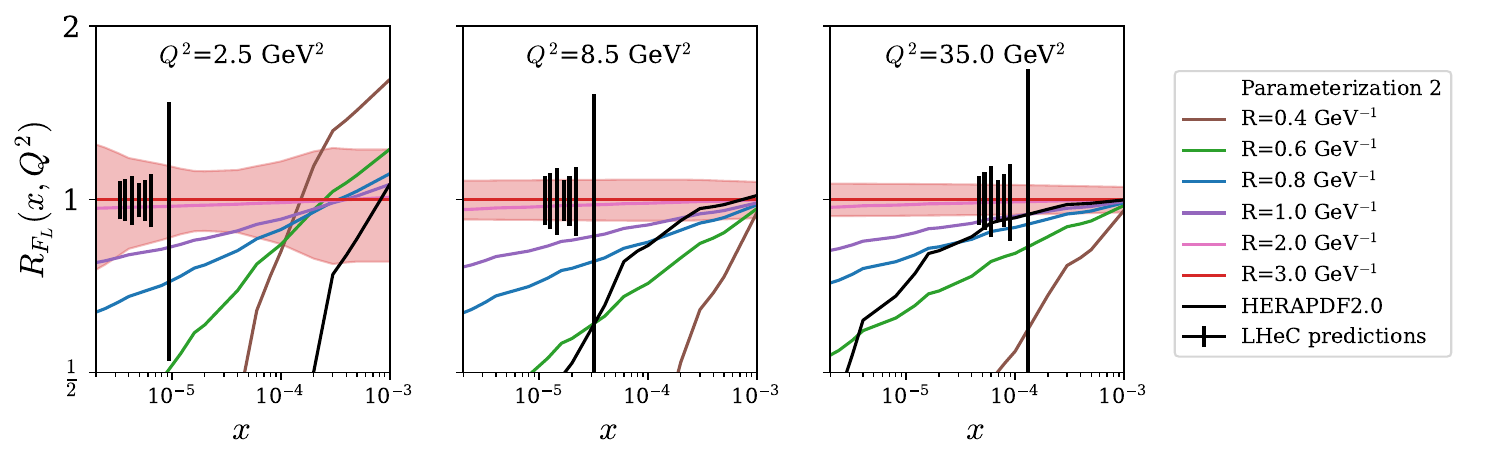}
    \caption{%
    The same as Fig.~\ref{fig:EIC_FL_2}, but for LHeC~\cite{LHeC:2020van}.
    }
    \label{fig:LHeC_FL_2}
\end{figure}

The observations made above naturally raise a question of how much better future DIS experiments such as EIC~\cite{Accardi:2012qut} and LHeC~\cite{LHeC:2020van, LHeCStudyGroup:2012zhm} could constrain the presence of non-linearities. 
To this end, we compare the spread of $F_L$ predictions by taking the ratio of the prediction resulting from each set of PDFs determined at a different value of $R$ over that at $R=3$\,GeV$^{-1}$:
\begin{align}
    R_{F_L}(x,Q^2)\equiv \frac{F_L(x,Q^2)}{F_L^{R=3\mathrm{GeV}^{-1}}(x,Q^2)}.\label{eqn:rfl}
\end{align}
These ratios are compared to the projected relative uncertainties for $F_L$ taken from the EIC development documents~\cite{Aschenauer:9999abc} and the LHeC White Paper~\cite{LHeC:2020van} in Figs.~\ref{fig:EIC_FL_2} and \ref{fig:LHeC_FL_2}, respectively, 
which are shown as black error bars. Again, we only show the results for parameterization 2 due to the negativity of $F_L$ when calculated with the PDFs from parameterization~1.
This 
allows us to 
estimate whether $F_L$ measurements at these future experiments can help to constrain the size of non-linear effects under the assumption that $F_L$ can be accurately reproduced by the theory, e.g., by adding small-$x$ resummation as in Refs.~\cite{Ball:2017otu, xFitterDevelopersTeam:2018hym}. The projected EIC data only reach the $x$ region, where the different $R$ values are clearly distinguishable at $Q<2.5$\,GeV$^2$, but even at $Q=2.47$\,GeV$^2$ the uncertainties of the pseudodata are barely smaller than the the spread between predictions with $R>0.6$\,GeV$^{-1}$. 
In the case of LHeC, however, the reach towards lower $x$ values should lead to more stringent constraints for the non-linear effects, since the different $R$ values lead to vastly different predictions in this kinematic region.

\section{Conclusions and Outlook}
\label{sec:conclusions}

We have presented numerical studies of the non-linear gluon recombination corrections to the DGLAP evolution as derived by Zhu and Ruan, which improve upon the commonly used GLR-MQ equations and restore the PDF momentum sum rule. We have extended the \textsc{HOPPET} and \textsc{xFitter} toolkits to account for these corrections. With these extensions we performed several fits of proton PDFs to BCDMS, HERA and NMC data on DIS at NNLO accuracy varying the dimensionful parameter $R$, which controls the strength of the recombination effects. We examined two different parametrizations for the gluon PDF -- one corresponding to HERAPDF2.0, which allows the gluon PDF to go negative at the parametrization scale, and another which is positive definite at the parametrization scale. We found that both parameterizations result in a similarly good
description of the data, $\chidof \sim 1.19$, for $R> 1$\,GeV$^{-1}$, but that the description begins to eventually deteriorate towards small $R$. For 
the 
parametrization allowing for a negative gluon PDF this deterioration sets in for $R \lesssim 0.7$\,GeV$^{-1}$, but for the positive-definite parametrization at somewhat lower values, $R \lesssim 0.5$\,GeV$^{-1}$. In both cases $R < 0.4$\,GeV$^{-1}$ seems to be excluded, which translates to an upper limit for the recombination scale $Q_r = 1/R \lesssim 2.5 \, {\rm GeV}$. We also studied the sensitivity of our fits to the choice of $Q^2_{\mathrm{cut}}$: lowering $Q^2_{\mathrm{cut}}$ from $3.50\,{\rm GeV}^2$ to $1 \,{\rm GeV}^2$ leads to an increase of $\chidof$, which
the non-linear corrections are unable to tame. In other words, we do not find evidence for the presence of non-linear effects in the used DIS data, but can set an upper limit for the parameter controlling their strength, $R \gtrsim 0.4 $\,GeV$^{-1}$ or $Q_r = 1/R \lesssim 2.5 \, {\rm GeV}$. 
Our conclusions about HERA data not supporting non-linear effects is in line with a recent study made in Ref.~\cite{Mantysaari:2018nng}.
We also compared our PDF fits with the longitudinal structure function $F_L$ available from HERA. While the ZEUS data lies below the predictions from our fits, the large uncertainties still make the two compatible. However, the H1 data, which reaches lower $Q^2$ values and has smaller uncertainties, lies well above the predictions at low $Q^2$. Non-linear effects in the evolution do not help to resolve this tension between NNLO calculations and the data, indicating that other theoretical ingredients, such as small-$x$ resummation are necessary to describe these data properly. Assuming that this tension can be resolved, we show that the improved statistics of the EIC, and particularly the reach towards lower $x$ values at the LHeC could provide significantly stronger constraints on non-linear effects. 

In general, after a short interval of $Q^2$ evolution, the non-linear effects tend to reduce both gluon and quark PDFs at small values of $x$ in comparison to the case with no recombination. This suppression becomes more pronounced as the parameter $R$ decreases and the effects can persist up to $Q\gtrsim100\,\mathrm{GeV}$. This behavior was not sensitive to the form of the initial parameterization and thereby appears to be a general feature of the non-linear effects. This is opposite to the BFKL resummation which tends to increase the small-$x$ PDFs at large $Q^2$. Nevertheless, the fact that the $Q^2$ evolution does not wash away all the possible remnants of non-linearities even at $Q={\cal O}(100\, \rm GeV)$ means that some effects could be seen in proton-proton collisions at the LHC as well. For example, the direct photon and heavy-flavor production at forward direction are sensitive to PDFs at small $x$ and some visible differences could be induced from the fits with different $R$. To chart such effects is one possible way to make use of the results of the present analysis. 
One can also extend the analysis itself by, e.g., including the contributions of two-gluon initiated processes in the coefficient functions as well. 
For example, the recombination contributions enter the coefficient functions of $F_L$ at $\mathcal{O}(\alpha_s^2)$, i.e., they are of the same perturbative order as the leading-twist NLO terms. As these leading-twist NLO terms turn out rather significant for $F_L$, the inclusion of higher-twist terms could potentially change the picture in the case of $F_L$. 
For $F_2$ the higher-twist terms are of the same perturbative order as the leading-twist NNLO terms. In addition, one 
could also include a wider range of data sets in the fit, e.g., from the LHC, and/or generalize the framework to the case of heavier nuclei with the expectation of observing enhanced effects of recombination. 
\\

The tools developed in this work and LHAPDF files~\cite{Buckley:2014ana} for the fitted PDFs are available here: \url{https://research.hip.fi/qcdtheory/nuclear-pdfs/}.

\section*{Acknowledgements}
Our work has been supported by the Academy of Finland, projects 331545 and 330448, and was funded as a part of the Center of Excellence in Quark Matter of the Academy of Finland, project 346326. This research is part of the European Research Council project ERC-2018-ADG-835105 YoctoLHC. We acknowledge grants of computer capacity from the Finnish Grid and Cloud Infrastructure (persistent identifier \mbox{urn:nbn:fi:research-infras-2016072533}).

\bibliographystyle{utphys}
\bibliography{refs.bib,extra.bib}
\printtables 
\printfigures

\appendix

\section{Obtained $\chi^2$ values for each data set}\label{sec:chi2data}

The following three Tables~\ref{tab:chi2table}-\ref{tab:chi2table_c} show the values of $\chidof$ obtained in the different sets of PDF fits for a selection of $R$ values $R=\{0.4, 0.6, 0.8, 1.0, 2.0, 3.0\}$\,GeV$^{-1}$. Table~\ref{tab:chi2table} shows the values obtained in the fits with parameterization 1 and $Q^2_{\mathrm{cut}}=3.5$\,GeV$^2$. Table~\ref{tab:chi2table_c} shows the same for $Q^2_{\mathrm{cut}}=1.0$\,GeV$^2$ and Table~\ref{tab:chi2table_b} for parameterization 2 with $Q^2_{\mathrm{cut}}=3.5$\,GeV$^2$.

\begin{table*}[htbp!]
    \renewcommand{\arraystretch}{1.4}
    \setlength\tabcolsep{4pt}
    \caption{Values of $\chidof$ obtained in the fits with parameterization 1 and $Q^2_{\mathrm{cut}} = 3.5$\,GeV$^2$ for various values for $R$ for each individual data set. The final row lists the total $\chidof$ for each $R$.}
    \centering	
    \begin{tabular}{|c|c|c|c|c|c|c|c|}
        \hline 
        \multicolumn{2}{|c|}{$R$ [GeV$^{-1}$]} & 0.4 & 0.6 & 0.8 & 1.0 & 2.0 & 3.0 \\ 
        \hline
		\hline
		\multirow{4}{*}{BCDMS}  & NC $\mu$ $F_2^p$ 100\,GeV & 1.257 & 1.273 & 1.236 & 1.229 & 1.202 & 1.197 \\ 
                                    & NC $\mu$ $F_2^p$ 120\,GeV & 0.770 & 0.766 & 0.753 & 0.754 & 0.748 & 0.746 \\ 
                                    & NC $\mu$ $F_2^p$ 200\,GeV & 1.118 & 1.106 & 1.091 & 1.093 & 1.087 & 1.086 \\ 
                                    & NC $\mu$ $F_2^p$ 280\,GeV & 0.949 & 0.947 & 0.946 & 0.951 & 0.948 & 0.948 \\ 
		\hline 
		
        \multirow{7}{*}{HERA}   & NC e$^+$ 920\,GeV & 1.324 & 1.248 & 1.218 & 1.210 & 1.200 & 1.200 \\ 
                                & NC e$^+$ 820\,GeV & 1.088 & 1.040 & 1.029 & 1.008 & 0.991 & 0.988 \\ 
                                & NC e$^+$ 575\,GeV & 0.924 & 0.895 & 0.883 & 0.879 & 0.875 & 0.874 \\ 
                                & NC e$^+$ 460\,GeV & 1.137 & 1.109 & 1.094 & 1.095 & 1.095 & 1.094 \\ 
                                & NC e$^-$ 920\,GeV & 1.495 & 1.469 & 1.457 & 1.458 & 1.454 & 1.455 \\ 
                                & CC e$^+$ 920\,GeV & 1.330 & 1.421 & 1.349 & 1.412 & 1.373 & 1.365 \\ 
                                & CC e$^-$ 920\,GeV & 1.505 & 1.434 & 1.501 & 1.505 & 1.494 & 1.491 \\ 
		\hline 
		NMC  & NC $\mu$ $F_2^p$ & 1.179 & 1.038 & 0.975 & 0.972 & 0.972 & 0.972 \\ 
		\hline 
        \hline
        \multicolumn{2}{|c|}{$\chidof$} & 1.244 & 1.205 & 1.194 & 1.188 & 1.182 & 1.182 \\
        \hline
    \end{tabular}     	
    \label{tab:chi2table}
\end{table*}

\begin{table*}[h]
    \renewcommand{\arraystretch}{1.4}
    \setlength\tabcolsep{4pt}
    \caption{The same as Table~\ref{tab:chi2table}, but with parameterization 2.}
    \centering	
    \begin{tabular}{|c|c|c|c|c|c|c|c|}
        \hline 
        \multicolumn{2}{|c|}{$R$ [GeV$^{-1}$]}                  & 0.4   & 0.6   & 0.8   & 1.0   & 2.0   & 3.0 \\ 
        \hline
		\hline
		\multirow{4}{*}{BCDMS}  & NC $\mu$ $F_2^p$ 100\,GeV & 1.045 & 1.006 & 1.012 & 1.013 & 1.016 & 1.017 \\ 
                                    & NC $\mu$ $F_2^p$ 120\,GeV & 0.755 & 0.749 & 0.749 & 0.750 & 0.751 & 0.752 \\ 
                                    & NC $\mu$ $F_2^p$ 200\,GeV & 1.090 & 1.083 & 1.082 & 1.082 & 1.082 & 1.082 \\ 
                                    & NC $\mu$ $F_2^p$ 280\,GeV & 0.934 & 0.934 & 0.935 & 0.935 & 0.937 & 0.937 \\ 
		\hline 
        \multirow{7}{*}{HERA}               & NC e$^+$ 920\,GeV & 1.211 & 1.188 & 1.188 & 1.191 & 1.197 & 1.198 \\ 
                                            & NC e$^+$ 820\,GeV & 1.066 & 1.041 & 1.029 & 1.024 & 1.021 & 1.021 \\ 
                                            & NC e$^+$ 575\,GeV & 0.883 & 0.878 & 0.877 & 0.877 & 0.878 & 0.878 \\ 
                                            & NC e$^+$ 460\,GeV & 1.074 & 1.066 & 1.068 & 1.070 & 1.072 & 1.073 \\ 
                                            & NC e$^-$ 920\,GeV & 1.455 & 1.455 & 1.456 & 1.456 & 1.457 & 1.457 \\ 
                                            & CC e$^+$ 920\,GeV & 1.006 & 0.952 & 0.946 & 0.944 & 0.942 & 0.941 \\ 
                                            & CC e$^-$ 920\,GeV & 1.218 & 1.263 & 1.262 & 1.261 & 1.262 & 1.262 \\ 
		\hline 
		NMC                              & NC $\mu$ $F_2^p$ & 1.092 & 1.082 & 1.081 & 1.083 & 1.086 & 1.087 \\ 
		\hline 
        \hline
        \multicolumn{2}{|c|}{$\chidof$} & 1.206 & 1.187 & 1.187 & 1.188 & 1.191 & 1.192 \\
        \hline
    \end{tabular}     	
    \label{tab:chi2table_b}
\end{table*}

\begin{table*}[htbp!]
    \renewcommand{\arraystretch}{1.4}
    \setlength\tabcolsep{4pt}
    \caption{The same as Table~\ref{tab:chi2table}, but with the lower cut $Q^2_{\mathrm{cut}} = 1.0$\,GeV$^2$. }
    \centering	
    \begin{tabular}{|c|c|c|c|c|c|c|c|}
        \hline 
        \multicolumn{2}{|c|}{$R$ [GeV$^{-1}$]} & 0.4 & 0.6 & 0.8 & 1.0 & 2.0 & 3.0 \\ 
        \hline
		\hline
		\multirow{4}{*}{BCDMS}  & NC $\mu$ $F_2^p$ 100\,GeV     & 1.272 & 1.262 & 1.227 & 1.206 & 1.175 & 1.167 \\ 
                                    & NC $\mu$ $F_2^p$ 120\,GeV     & 0.797 & 0.762 & 0.755 & 0.751 & 0.747 & 0.746 \\ 
                                    & NC $\mu$ $F_2^p$ 200\,GeV     & 1.133 & 1.099 & 1.092 & 1.089 & 1.087 & 1.087 \\ 
                                    & NC $\mu$ $F_2^p$ 280\,GeV     & 0.955 & 0.949 & 0.952 & 0.952 & 0.952 & 0.952 \\ 
		\hline 
		
        \multirow{7}{*}{HERA}   & NC e$^+$ 920\,GeV & 1.778 & 1.331 & 1.256 & 1.231 & 1.213 & 1.212 \\ 
                                & NC e$^+$ 820\,GeV & 1.816 & 1.218 & 1.213 & 1.196 & 1.179 & 1.177 \\ 
                                & NC e$^+$ 575\,GeV & 1.017 & 0.919 & 0.907 & 0.901 & 0.898 & 0.898 \\ 
                                & NC e$^+$ 460\,GeV & 1.229 & 1.111 & 1.078 & 1.081 & 1.076 & 1.075 \\ 
                                & NC e$^-$ 920\,GeV & 1.560 & 1.495 & 1.472 & 1.466 & 1.463 & 1.463 \\ 
                                & CC e$^+$ 920\,GeV & 1.420 & 1.332 & 1.362 & 1.361 & 1.363 & 1.361 \\ 
                                & CC e$^-$ 920\,GeV & 2.039 & 1.598 & 1.504 & 1.494 & 1.484 & 1.480 \\ 
		\hline 
		NMC  & NC $\mu$ $F_2^p$ & 1.648 & 1.068 & 1.008 & 1.008 & 1.023 & 1.028 \\ 
		\hline 
        \hline
        \multicolumn{2}{|c|}{$\chidof$} & 1.507 & 1.259 & 1.226 & 1.218 & 1.213 & 1.212 \\
        \hline
    \end{tabular}     	
    \label{tab:chi2table_c}
\end{table*}

\FloatBarrier
\section{Obtained PDF parameters}\label{sec:pdfparams}
The following three Tables~\ref{tab:paramtable}-\ref{tab:paramtable_c} show the parameters obtained in the different sets of PDF fits for a selection of $R$ values $R=\{0.4, 0.6, 0.8, 1.0, 2.0, 3.0\}$\,GeV$^{-1}$. Table~\ref{tab:paramtable} shows the parameters obtained in the fits with parameterization 1 and $Q^2_{\mathrm{cut}}=3.5$\,GeV$^2$. Table~\ref{tab:paramtable_b} shows the same for $Q^2_{\mathrm{cut}}=1.0$\,GeV$^2$ and Table~\ref{tab:paramtable_c} for parameterization 2 with $Q^2_{\mathrm{cut}}=3.5$\,GeV$^2$.
\begin{table*}[htbp!]
    \renewcommand{\arraystretch}{1.4}
    \setlength\tabcolsep{4pt}
    \caption{Parameters obtained in the fits with parameterization 1 and $Q^2_{\mathrm{cut}} = 3.5$\,GeV$^2$ at various values for $R$. Parameters in parentheses are fixed at the given value, while parameters not shown are set to zero.}
    \centering	
    \begin{tabular}{|c|c|c|c|c|c|c|}
        \hline 
        $R$ [GeV$^{-1}$] & 0.4 & 0.6 & 0.8 & 1.0 & 2.0 & 3.0 \\ 
        \hline
        \hline
          $A_{\bar{d}}$ & 0.404469 & 0.161368 & 0.104539 & 0.061467 & 0.056025 & 0.054686 \\ 
          $A'_{g}$ & 2.218085 & 0.435239 & 0.410244 & 0.201921 & 0.494136 & 0.428424 \\ 
          $B_{\bar{d}}$ & 0.010765 & -0.085343 & -0.108657 & -0.128949 & -0.144103 & -0.145589 \\ 
          $B_{d_v}$ & 1.422865 & 1.454721 & 1.363376 & 1.323526 & 1.314441 & 1.313151 \\ 
          $B_{g}$ & -0.211063 & -0.342516 & -0.174697 & -0.042353 & -0.123857 & -0.069253 \\ 
          $B'_{g}$ & 0.256574 & -0.449499 & -0.309159 & -0.302806 & -0.260254 & -0.235 \\ 
          $B_{u_v}$ & 0.914925 & 0.921869 & 0.934284 & 0.921569 & 0.923402 & 0.923452 \\ 
          $C_{\bar{d}}$ & 18.358827 & 13.091428 & 6.798707 & 2.759383 & 2.15534 & 2.020401 \\ 
          $C_{d_v}$ & 5.876824 & 5.987926 & 5.560861 & 5.694443 & 5.733379 & 5.735318 \\ 
          $C_{g}$ & 2.208999 & 2.120541 & 3.683178 & 4.354789 & 4.851641 & 5.230455 \\ 
          $C'_{g}$ & (25.0) & (25.0) & (25.0) & (25.0) & (25.0) & (25.0) \\ 
          $C_{\bar{u}}$ & 15.873966 & 17.198562 & 16.234897 & 16.742077 & 15.787451 & 15.521866 \\ 
          $C_{u_v}$ & 2.48766 & 2.496799 & 2.529641 & 2.489838 & 2.489735 & 2.487229 \\ 
          $D_{\bar{u}}$ & 16.866158 & 28.489481 & 22.125291 & 23.503082 & 22.000836 & 21.205419 \\ 
          $E_{u_v}$ & -0.997742 & -0.988268 & -0.971517 & -1.003805 & -1.020271 & -1.027135 \\ 
        \hline 
    \end{tabular}     	
    \label{tab:paramtable}
\end{table*}

\begin{table*}[htbp!]
    \renewcommand{\arraystretch}{1.4}
    \setlength\tabcolsep{4pt}
    \caption{The same as Table~\ref{tab:paramtable}, but for PDFs fitted with the lower cut $Q^2_{\mathrm{cut}}=1.0$\,GeV$^2$.}
    \centering	
    \begin{tabular}{|c|c|c|c|c|c|c|}
        \hline 
        $R$ [GeV$^{-1}$] & 0.4 & 0.6 & 0.8 & 1.0 & 2.0 & 3.0 \\ 
        \hline
        \hline
          $A_{\bar{d}}$ & 0.150472 & 0.105445 & 0.094215 & 0.082058 & 0.06272 & 0.061345 \\ 
          $A'_{g}$ & -0.481786 & 0.104096 & 0.142766 & 0.234982 & 0.24655 & 0.268521 \\ 
          $B_{\bar{d}}$ & -0.089444 & -0.091097 & -0.100384 & -0.104039 & -0.109834 & -0.110594 \\ 
          $B_{d_v}$ & 1.026321 & 1.222286 & 1.314041 & 1.297563 & 1.266633 & 1.265011 \\ 
          $B_{g}$ & 0.112897 & -0.090972 & 0.012384 & 0.019837 & 0.103054 & 0.1072 \\ 
          $B'_{g}$ & -0.148734 & -0.378519 & -0.278353 & -0.215802 & -0.153492 & -0.140128 \\ 
          $B_{u_v}$ & 0.872954 & 0.913072 & 0.927043 & 0.924927 & 0.919602 & 0.91878 \\ 
          $C_{\bar{d}}$ & 4.207006 & 5.524392 & 6.20659 & 5.084081 & 2.925174 & 2.766838 \\ 
          $C_{d_v}$ & 4.871045 & 5.154293 & 5.408388 & 5.359225 & 5.395968 & 5.409781 \\ 
          $C_{g}$ & 2.001991 & 2.939454 & 4.006964 & 4.490556 & 5.487986 & 5.642055 \\ 
          $C'_{g}$ & (25.0) & (25.0) & (25.0) & (25.0) & (25.0) & (25.0) \\ 
          $C_{\bar{u}}$ & 22.296374 & 16.795233 & 15.487592 & 14.595522 & 13.624455 & 13.376877 \\ 
          $C_{u_v}$ & 2.365049 & 2.476635 & 2.499119 & 2.481032 & 2.461876 & 2.458407 \\ 
          $D_{\bar{u}}$ & 30.234495 & 17.061131 & 15.205939 & 12.889824 & 10.462935 & 9.919356 \\ 
          $E_{u_v}$ & -1.056546 & -1.003855 & -0.993606 & -1.007731 & -1.02491 & -1.028614 \\ 
        \hline 
    \end{tabular}     	
    \label{tab:paramtable_b}
\end{table*}

\begin{table*}[htbp!]
    \renewcommand{\arraystretch}{1.4}
    \setlength\tabcolsep{4pt}
    \caption{The same as Table~\ref{tab:paramtable}, but for PDFs fitted with parameterization 2.}
    \centering	
    \begin{tabular}{|c|c|c|c|c|c|c|}
        \hline 
        $R$ [GeV$^{-1}$] & 0.4 & 0.6 & 0.8 & 1.0 & 2.0 & 3.0 \\ 
        \hline
        \hline
          $A_{\bar{d}}$ & 0.248371 & 0.238772 & 0.229188 & 0.226517 & 0.224377 & 0.224177 \\ 
          $B_{\bar{d}}$ & -0.095279 & -0.108632 & -0.115726 & -0.118081 & -0.120341 & -0.120637 \\ 
          $B_{d_v}$ & 0.989654 & 1.05047 & 1.050583 & 1.053705 & 1.057556 & 1.058181 \\ 
          $B_{g}$ & -0.063721 & 0.176504 & 0.300771 & 0.368229 & 0.46931 & 0.489468 \\ 
          $B_{u_v}$ & 0.330037 & 0.10342 & 0.10933 & 0.108711 & 0.108755 & 0.108843 \\ 
          $C_{\bar{d}}$ & 9.978846 & 10.592777 & 10.523469 & 10.595957 & 10.745633 & 10.781037 \\ 
          $C_{d_v}$ & 4.363475 & 4.575691 & 4.5757 & 4.591326 & 4.61234 & 4.61603 \\ 
          $C_{g}$ & 19.131659 & 21.235982 & 21.791519 & 21.993852 & 22.228964 & 22.267116 \\ 
          $C_{\bar{u}}$ & 8.828193 & 9.34704 & 9.525786 & 9.554065 & 9.563123 & 9.560328 \\ 
          $C_{u_v}$ & 3.24987 & 2.501855 & 3.573471 & 3.572397 & 3.568641 & 3.567745 \\ 
          $D_{g}$ & -9.671812 & -8.386368 & -7.830814 & -7.681217 & -7.689173 & -7.720356 \\ 
          $D_{\bar{u}}$ & 9.759016 & 10.223122 & 11.255959 & 11.466205 & 11.578213 & 11.574223 \\ 
          $D_{u_v}$ & 31.344142 & 232.987381 & 215.930099 & 217.666728 & 217.330601 & 216.978654 \\ 
          $E_{g}$ & 465.912581 & 399.008994 & 335.978167 & 302.820564 & 257.896509 & 249.69638 \\ 
          $E_{u_v}$ & -20.809607 & -238.165353 & 11.060703 & 11.049585 & 9.585398 & 9.200095 \\ 
        \hline 
    \end{tabular}     	
    \label{tab:paramtable_c}
\end{table*}

\end{document}